\documentclass[twocolumn,superscriptaddress,aps,preprintnumbers,amsmath,amssymb,prl,nofootinbib,10pt]{revtex4-2}

\usepackage{graphicx} 
\usepackage{dcolumn}
\usepackage{bm}
\usepackage[T1]{fontenc}
\usepackage{booktabs}
\usepackage{bbold}
\usepackage{mathtools}
\usepackage{multirow}
\usepackage{physics}
\usepackage{slashed}
\usepackage{colortbl}
\usepackage{hyperref}
\usepackage[dvipsnames]{xcolor}
\usepackage{orcidlink}
\usepackage{tikz,xcolor,hyperref}
\usepackage{adjustbox}
\usepackage{cancel}

\graphicspath{{Figures/}}
\usepackage[compat=1.1.0]{tikz-feynman}

\newcommand\myshade{85}
\colorlet{mylinkcolor}{Blue}
\colorlet{mycitecolor}{Blue}
\colorlet{myurlcolor}{NavyBlue}
\hypersetup{
     linkcolor  = mylinkcolor,
     citecolor  = mycitecolor!\myshade!black,
     urlcolor   = myurlcolor!\myshade!black,
     colorlinks = true,
}

\definecolor{Gray}{gray}{0.85}
\definecolor{LightCyan}{rgb}{0.88,1,1}
\newcolumntype{a}{>{\columncolor{Gray}}c}

\DeclareMathOperator{\eV}{\text{eV}}

\DeclareMathOperator{\GeV}{\text{GeV}}

\begin{document}

\title{Dark Energy and Neutrino Flavor from the Weak Axion}

\author{Pedro Bittar \orcidlink{0000-0002-3684-5692}}
\email{bittar.hep@gmail.com}
\affiliation{Perimeter Institute for Theoretical Physics, Waterloo, ON, N2L 2Y5, Canada}

\author{Carlos E. M. Wagner \orcidlink{0000-0001-6407-623X}}
\email{cwagner@uchicago.edu}
\affiliation{Perimeter Institute for Theoretical Physics, Waterloo, ON, N2L 2Y5, Canada}
\affiliation{HEP Division, Argonne National Laboratory, 9700 Cass Ave., Argonne, IL 60439, USA}
\affiliation{Enrico Fermi Institute, Physics Department, University of Chicago, Chicago, IL 60637, USA}
\affiliation{Kavli Institute for Cosmological Physics, University of Chicago, Chicago, IL 60637, USA}
\affiliation{Leinweber Institute for Theoretical Physics, University of Chicago, Chicago, IL 60637, USA} 

\date{\today}

\begin{abstract}
\noindent Dynamical dark energy offers an alternative to a cosmological constant with distinct observational signatures. However, the small energy density scale, Hubble-sized mass, and Planckian excursions make simple models fine-tuned and unnatural. In this work, we show that a weak version of the axion, identified with the phase field of the anomalous $U(1)_{B+L}$ of the Standard Model, can generate the scale hierarchies expected for dark energy. The axion potential is controlled by sources of explicit baryon and lepton number violation and is radiatively stable. We show that the leading contribution comes from two inequivalent Weinberg operators, one $B+L$-conserving and one $B+L$-violating, which generate the axion potential. We propose a flavor selection rule in the lepton sector that simultaneously removes the quadratic divergence and dominant temperature-dependent contributions to the effective potential. The resulting potential first appears at quartic order in the axion-dependent neutrino masses and, depending on the flavor model and PMNS parameters, its amplitude is parametrically close to the dark-energy density. Since the scale of dark energy is sensitive to changes in the neutrino parameters, improved measurements of  $m_{\beta\beta}$, the sum of neutrino masses, $\delta_{\rm CP}$ and $\theta_{23}$ can favor certain models. Cosmologically, the field behaves as thawing quintessence, stays frozen by Hubble friction until late times and never enters the adiabatic regime.
\end{abstract}

\maketitle

\section{Introduction}
\label{sec:I}
In the last decades, observations from cosmological distances~\cite{SupernovaSearchTeam:1998fmf,SupernovaCosmologyProject:1998vns,Moresco:2016mzx,eBOSS:2020yzd}, the cosmic microwave background (CMB) \cite{Planck:2018vyg, Sherwin:2011gv}, and growth of structure \cite{Nadathur:2020kvq,Rose:2020shp} show that the universe is in a phase of accelerated expansion. This expansion reveals the presence of a dominating fraction of dark energy, a component of the universe with a negative effective equation of state. A small cosmological constant (CC) is the simplest explanation for dark energy. The value of the CC is the sum of a bare gravitational CC and the contributions to the vacuum energy from quantum fields. However, for an electroweak cutoff, vacuum energy is estimated to be  $55$ orders of magnitude above the observed value for the CC of $\Lambda^4 \simeq (2.3 \times 10^{-3} \eV)^4$~\cite{Planck:2018vyg}. For this reason, the CC leads to the most severe naturalness problem our current model of the universe faces~\cite{Weinberg:1988cp,Peebles:2002gy}, requiring enormous fine-tuning to fit the observed value.  

Although cosmic acceleration and vacuum energy are usually linked, it is possible that dark energy is not dominated by non-dynamical vacuum energy contributions but instead by the potential energy of new dynamical fields. In quintessence models \cite{Copeland:2006wr}, dark energy is driven by a nearly massless, slowly rolling scalar field. The main motivation for dynamical dark energy is that it offers more observational tests than a single-parameter CC. For example, a time-dependent equation of state $w(z)$ affects the late-time expansion history and the growth of structure, and can be probed by different cosmological observations \cite{SupernovaSearchTeam:2004lze,SDSS:2005xqv,Jimenez:2001gg,Crittenden:1995ak,Planck:2018vyg,Guzzo:2008ac,DES:2017myr,Vikhlinin:2008ym}. If the quintessence field has additional couplings to matter or radiation, it may also lead to fifth forces, birefringence, or time variation of fundamental constants \cite{Khoury:2003aq,Carroll:1998zi,Uzan:2002vq}. Additionally, recent DESI and DES measurements have renewed interest in such departures from a cosmological constant \cite{DESI:2024mwx,DESI:2025zgx,DESI:2024aqx,DESI:2025fii,DES:2026aht}.

Despite the observational appeal, dynamical dark energy faces several conceptual and phenomenological challenges. First, it relies on a separate solution to the CC problem, as it does not explain how the vacuum energy could decouple from gravity. Additionally, models of scalar field dark energy must generate a potential that leads to slow-roll evolution today. This requires an energy density $\rho_{\rm DE} \simeq (2.3 \times 10^{-3} \mathrm{eV})^4$, fixed by the observed expansion rate, and a mass $m_\phi \lesssim H_0 \sim 10^{-33}\,\mathrm{eV}$ so that the field evolves on Hubble timescales. In addition, the slow-roll condition $M_{\rm pl}\,|V'|/V \lesssim 1$ usually implies Planck scale field excursions. These extreme scale hierarchies pose an obstacle to model building. If the field couples to matter, its lightness generically leads to long-range forces that are tightly constrained by fifth-force searches~\cite{Carroll:1998zi,Khoury:2003aq}. Moreover, the large hierarchy between the dark energy scale and the other Standard Model (SM) scales renders the potential highly sensitive to radiative corrections, which can easily spoil the flatness required for cosmic acceleration. As a result, many simple scalar dark energy models are severely fine-tuned \cite{Kolda:1998wq}.

Several approaches have been proposed to address these issues. Attempts to \textit{relax} \cite{Abbott:1984qf,Brown:1987dd,Brown:1988kg,Graham:2019bfu,Bousso:2000xa,Polchinski:2006gy}, \textit{forbid} \cite{Obied:2018sgi,Ooguri:2018wrx,Dvali:2014gua,Dvali:2013eja,Dvali:2020etd} or \textit{sequester} \cite{Kaloper:2014dqa,Kaloper:2015jra,Kaloper:2016yfa,Kaloper:2016jsd,DAmico:2017ngr,Coltman:2019mql,Khoury:2026eqr} the vacuum energy\footnote{Among these, relaxation mechanisms aim to dynamically drive the vacuum energy toward a small or vanishing value, while other proposals are often tied to the conjecture that gravity forbids stable de Sitter vacua. Vacuum energy sequestering imposes global constraints to remove the radiative contributions from the matter vacuum energy from the gravitational field equations. While no consensus has yet emerged, these ideas have clarified possible mechanisms and their limitations.}, motivate a possibly zero CC. Ultraviolet quintessence constructions aimed at stabilizing the large hierarchy of scales \cite{Binetruy:1998rz,Brax:1999gp,Brax:1999yv,Copeland:2000vh,Frieman:1995pm,Choi:1999xn,Kim:2002tq,Gasperini:2001pc,Damour:2002mi,Burgess:2021obw} and mechanisms designed to avoid strong phenomenological constraints \cite{Khoury:2003rn,Khoury:2003aq,Brax:2004px,Hinterbichler:2010es} were also proposed. In this paper, we will not attempt to solve the CC problem and will assume that the vacuum energy is set to zero, while remaining agnostic about the underlying mechanism responsible for this. Instead, we focus on a theoretically well-motivated and phenomenologically viable candidate for dark energy, the weak axion.

We define the weak axion as the field whose shift symmetry is associated with the anomalous $U(1)_{B+L}$ global symmetry of the SM. This axion has been considered in different contexts in the literature before \cite{McLerran:2012mm,Nomura:2000yk,Shifman:2017lkj,Dvali:2024zpc,Dvali:2025pcx,Cacciapaglia:2025xmr,Cacciapaglia:2025dme,Davoudiasl:2025qqv,Berbig:2025nrt}. Unlike the QCD axion, where strong instantons generate a mass set by the QCD scale, electroweak instantons lead to a mass exponentially suppressed with respect to the weak scale once the Higgs acquires a vacuum expectation value (VEV). As a result, if there are no other explicit sources of $B+L$ breaking, the weak axion mass is naturally extremely small. We revisit the calculations of the weak axion potential arising from instanton effects \cite{Nomura:2000yk,McLerran:2012mm} and conclude that, in our setup, they are below the scale required for dark energy. 

Since $B+L$ violating terms are accidentally suppressed in the SM, they provide controlled sources of explicit symmetry breaking and a radiatively stable axion potential. Furthermore, due to the proximity between dark energy density scale and neutrino masses, a dark energy sized potential  with the right magnitude arises  if the dominant explicit breaking is tied to an axion varying Majorana neutrino mass. This connection was explored in mass-varying neutrino models \cite{Fardon:2003eh,Peccei:2004sz} more than two decades ago, but those models relied on strong adiabatic evolution of the scalar field, leading to instabilities and the formation of neutrino non-linear overdensities, i.e. neutrino nuggets \cite{Afshordi:2005ym}. 

In the case of the weak axion, we show that the simplest realization of the model behaves like standard thawing quintessence at the background level and is free of the instabilities present in mass-varying neutrino models, since the field does not follow the adiabatic regime. Moreover, we impose a flavor selection rule that removes the one-loop quadratic divergence and the leading temperature-dependent contribution. Requiring the two neutrino mass terms to have no overlapping nonzero entries in the charged-lepton mass basis also removes the quadratic invariant with charged-lepton insertions. We focus on electron- and muon-aligned textures satisfying this condition. Their neutrino-induced potential is fixed once the flavor direction, neutrino masses, mixing parameters, and Majorana phases are specified. Improved neutrino measurements can therefore sharpen the connection to the dark-energy scale.

The paper is organized as follows. In Section \ref{sec:II}, we define the axion lagrangian and the low energy effective theory. In Section \ref{sec:III}, we discuss the cosmological dynamics of the relics. In Section \ref{sec:V}, we present our conclusions. 

\section{The Weak Axion}
\label{sec:II}

Light scalar fields protected by shift symmetries are natural candidates for dark energy provided that the associated explicit breaking is extraordinarily small. Baryon and lepton number symmetries automatically arise as accidental global symmetries of very high quality in the SM. These are broken only by the electroweak anomaly and possibly by neutrino masses and explicit baryon number violation in BSM models. Such controlled sources of explicit breaking make $B+L$ a natural choice for the symmetry associated with a very light weak axion. 

We define the weak axion as the phase of a complex scalar field $\Phi$ that carries a $B+L$ global charge,
\begin{align}
    \Phi(x) = \frac{f}{\sqrt{2}} e^{-i a(x)/{f}},
\end{align}

\noindent where $f$ denotes the axion decay constant. To connect this field with baryon and lepton number, we can write the following effective lagrangian
\begin{align}
    \nonumber \mathcal{L} =& \,|\partial_\mu \Phi|^2 + \frac{C_\nu}{f^2} LHLH \Phi + \frac{C_{BL}}{f^3} QQQL \Phi + h.c.
    \\
    & \hspace{0.7cm} + \frac{g^2}{32\pi^2} \theta_W W_{\mu\nu}^a \widetilde W^{\mu\nu\, a} + \mathcal{L}_{BLV}.
    \label{eq:lag}
\end{align}

\noindent Note that the presence of the electroweak vacuum angle $\theta_W$ is physical since we allow for explicit $B+L$ violation through the operators in $\mathcal{L}_{BLV}$. Neglecting the anomaly and assuming that the explicit breaking operators are parametrically smaller than the other terms in Eq.\eqref{eq:lag}, we can assign the global $B+L$ charges to the fields according to table \ref{tab:B+L}.

\begin{table}
    \centering
    \renewcommand{\arraystretch}{1.4}
    \setlength{\tabcolsep}{6pt}
    \begin{tabular}{|c|cccccccc|}
        \hline
              & $Q$ & $u^c$ & $d^c$ & $L$ & $e^c$ & $\nu^c$ & $H$ & $\Phi$ \\
        \hline
        $B+L$ & $\tfrac{1}{3}$
              & $-\tfrac{1}{3}$
              & $-\tfrac{1}{3}$
              & $1$
              & $-1$
              & $-1$
              & $0$
              & $-2$ \\
        \hline
    \end{tabular}
    \caption{Approximate $B+L$ global charge assignments.}
     \label{tab:B+L}
\end{table}

We can eliminate the weak axion from the $B+L$ preserving terms with the following field redefinitions,
\begin{alignat}{5}
    &L \mapsto e^{ia/(2f)} L, \qquad && Q \mapsto e^{ia/(6f)} Q,
    \label{eq:LH_transf}
    \\
    &e^c \mapsto e^{-ia/(2f)} e^c, \qquad && q^c \mapsto e^{-ia/(6f)} q^c,
    \label{eq:RH_transf}
\end{alignat}

\noindent where $q^c\equiv u^c,d^c$. This $B+L$ transformation induces a $(\partial^\mu a)J^\mu_{B+L}$ interaction and couplings of the axion to the weak, hypercharge, and gravitational topological terms through the mixed anomalies.
\begin{align}
&\mathcal{L}' =
    \frac{1}{2}(\partial_\mu a)^2
    + \frac{\partial_\mu a}{f} J_{B+L}^\mu
    + \frac{g^2}{32\pi^2}
      \left(\theta_W + N_f\frac{a}{f}\right)
      W_{\mu\nu}^a \widetilde W^{\mu\nu a}
\notag \\
&
    + \left(
        \frac{\sqrt{2}C_\nu}{f} LHLH
        + \frac{C_{BL}}{\sqrt{2}f^2} QQQL
        + \mathrm{h.c.}
      \right)
    + \mathcal{L}'_{\rm BLV}(a)
\label{eq:lagaxion}
\end{align}

\noindent where we omitted the term generated by the hypercharge and gravitational anomaly\footnote{The gravitational anomaly is only generated if the number of right handed neutrinos is different from three.}. The $B+L$ current is defined with the fermionic fields and their charges as defined in table \ref{tab:B+L}. Throughout this article, we shall use the form for the lagrangian Eq.\eqref{eq:lag} or Eq.\eqref{eq:lagaxion} that is most convenient given the specific calculation at hand.

If the weak axion is to account for dark energy, the potential must reproduce the observed dark energy density today and the mass of the axion must be of order of the Hubble constant. The relevant scales are
\begin{gather}
    V(a_0) \simeq 3 \,\Omega_{\rm DE} \,H_0^2 \, M_{\rm pl}^2 \simeq (2.3 \times 10^{-3} \, \text{eV})^4
    \\[1.5mm]
    m_a^2 \mathrel \sim H_0^2 \simeq (1.5 \times 10^{-33} \eV)^2
    \\
    f^2 \simeq \frac{V(a_0)}{m_a^2} \sim (10^{27} \eV)^2
\end{gather}
 For a generic quintessence field, this huge scale hierarchy is difficult to stabilize. In the case of the weak axion, we should carefully identify the sources of explicit $B+L$ breaking that can realistically set the potential.

\subsection{Weak Instanton potential}

The first candidate for explicit breaking is the electroweak anomaly. The possibility that weak instantons generate a dark energy--sized potential has been explored in the literature \cite{McLerran:2012mm,Nomura:1999py,Nomura:2000yk}. However,  assuming the SM running of the weak coupling, the resulting contribution is below the scale required for dark energy, as discussed next.

Instantons are semiclassical field configurations that interpolate between vacua with different Chern--Simons number and carry topological charge, $Q \in \mathbb{Z}$. 
Since the axion couples to the topological density, a slowly varying axion
background contributes as a phase to the Euclidean action,
\begin{align}
    \int d^4x\,
    \left(
        \frac{a}{f}+\theta_W
    \right)
    \frac{g^2}{32\pi^2}
    W_{\mu\nu}\widetilde W^{\mu\nu}
    =
    \left(
        N_f\frac{a}{f}+\theta_W
    \right)Q ,
\end{align}

\noindent The constrained-instanton action is
\begin{equation}
    S_E(\rho)
    =
    \frac{8\pi^2}{g^2(1/\rho)}
    +\pi^2v^2\rho^2
    +\mathcal{O}(\lambda\rho^4v^4).
    \label{eq:SEH}
\end{equation}

\noindent Therefore, the Higgs VEV breaks scale invariance and exponentially suppresses instantons with $\rho\gg (gv)^{-1}$. 

In addition to the Higgs suppression from the instanton action, a non-vanishing contribution to the potential requires saturating the twelve fermionic zero modes implied by the index theorem. These can be saturated by three insertions of the $QQQL$ operator from Eq.~\eqref{eq:lagaxion}. Writing $ c_{BL}=|c_{BL}|e^{i\delta_{BL}}$,
the resulting potential takes the form
\begin{align}
    V_{\rm inst}(a)
    =
    -\Lambda_{\rm inst}^4
    \cos\left(
        N_f\frac{a}{f}
        +\theta_W
        +3\delta_{BL}
    \right).
\end{align}

\noindent Following Refs.~\cite{Morrissey:2005uza,Csaki:2023ziz}, the weak-instanton amplitude is
\begin{align}
    &\Lambda_{\rm inst}^4
    \simeq{}
    2C_2
    \left(
        \frac{8\pi^2}{g^2(M_{UV})}
    \right)^4
    \frac{|c_{BL}|^3}{(4\pi)^6}
    ~\mathcal{I}[M_{UV}],
    \label{eq:weak-instanton-amplitude}
    \\
    & \mathcal{I}[M_{UV}]\equiv \int_{M_{UV}^{-1}}^\infty \frac{d\rho}{\rho^5} (\Lambda_{SU(2)} \rho)^{b_0}\frac{1}{(\rho f)^6} e^{-\pi^2 \rho^2 v^2}
\end{align}

\noindent For more details on the instanton potential calculation we refer to App.\ref{app:instanton}.

Assuming the SM weak beta function, the instanton-size integral is cut off at $\rho\simeq M_{\rm UV}^{-1}$. Then, taking $M_{UV} \simeq f \simeq M_{\rm pl}$, the numerical estimate gives
\begin{align}
    \Lambda_{\rm inst}^4
    \simeq{}&
    \left(6\times10^{-6}\,\mathrm{eV}\right)^4
    |c_{BL}|^3
\end{align}

\noindent Therefore, the potential is about $10$ orders of magnitude too small to account for dark energy. This estimate relies on the SM running of the weak gauge coupling up to $M_{\rm UV}$ and on the saturation of the fermionic zero modes using the operators in \eqref{eq:lagaxion}. Once the electroweak zero modes are saturated, the instanton size integral is sensitive to the UV completion.  This observation underlies earlier weak axion dark energy constructions, including the supersymmetric setup of Ref.~\cite{Nomura:2000yk} and the small-instanton analysis of Ref.~\cite{McLerran:2012mm}. In particular, new degrees of freedom can increase $\alpha_2(M_{\rm UV})$ and enhance the instanton potential exponentially. However, such an enhancement depends on the UV spectrum and on the detailed structure of the $B+L$-violating operators.  We will not attempt to engineer this possibility here.  Instead, we take the estimate above as the minimal SM-running instanton contribution and focus below on a neutrino-related source of explicit $B+L$ violation, which dominates over the instanton-induced contribution in the parameter range of interest.

\subsection{Neutrino induced potential}

There are many possible sources for explicit terms that violate $B+L$. A minimal possibility is to supplement the dressed Weinberg operator in Eq.\eqref{eq:lag} by a second $LHLH$ term that does not share the same axion dependence,
\begin{equation}
    \mathcal{L}_{BLV} = \frac{C_\star}{\Lambda_\star} LHLH + h.c.
\end{equation}

\noindent We assume that this term is generated by some UV source at the scale $\Lambda_\star/C_\star^{ij}$. In the basis suggested by Eq.\eqref{eq:lag}, after electroweak symmetry breaking, the two Weinberg operators combine into the following Majorana mass for the active neutrinos,
\begin{align}
    &\mathcal{L}_\nu = -\frac{1}{2}\nu \,M_\nu(a) \, \nu -\frac{1}{2}\nu^\dagger \,M^\dagger_\nu(a) \, \nu^\dagger
\end{align}

\noindent The Majorana mass matrix is split into axion-dependent and axion-independent terms
\begin{align}
    & M_\nu(a) = M_{\Phi} \,e^{-i a/f} + M_\star,
    \label{eq:Mnua}
    \\
    & M_\Phi \equiv \frac{v^2}{f}C_\nu, \quad M_\star \equiv \frac{v^2}{\Lambda_\star}C_\star.
\end{align}

Due to the rolling of the axion, the active neutrino masses and mixing parameters are dynamical on cosmological timescales. The present-day neutrino parameters define the Majorana mass matrix
\begin{align}
    M_{\nu,0} \equiv M_\nu(a_0)
    =
    V^*
    \operatorname{diag}(m_1, m_2, m_3)
    V^\dagger ,
    \label{eq:Mnu_PMNS}
\end{align}
where $a_0$ is the present-day field value of the axion, $m_{1},m_2,m_3$ are the neutrino masses and $V$ is the PMNS matrix, which encodes the mixing angles $\theta_{12}$, $\theta_{13}$, $\theta_{23}$, the Dirac phase $\delta_{CP}$, and the two Majorana phases $\alpha_{21}$, $\alpha_{31}$. However, this present-day matrix does not determine the full field dependence of $M_\nu(a)$. That dependence is controlled by the flavor decomposition between $M_\Phi$ and $M_\star$, subject to the requirement that 
\begin{equation}
    M_{\nu,0} =M_\Phi e^{-ia_0/f}+M_\star
\end{equation}
reproduces the observed neutrino data today. Since this decomposition also determines how the axion enters the Coleman-Weinberg potential, additional flavor assumptions are needed to make the model predictive. As we discuss next, radiative stability imposes a further constraint on the allowed flavor structures.

Now, our goal is to generate the Coleman-Weinberg potential from the masses in Eq.\eqref{eq:Mnua}. However, the potential has an undesired quadratically divergent sensitivity to the cutoff in general. This contribution comes from the trace of the Majorana mass squared,
\begin{align}
    \frac{\Lambda_{\rm UV}^2}{16\pi^2} \mathrm{Tr}\big[M_\nu^\dagger M_\nu\big] = \frac{\Lambda_{\rm UV}^2}{16\pi^2} e^{ia/f} \,\mathrm{Tr}\big[M_\Phi^\dagger M_\star\big] + h.c.
    \label{eq:HP}
\end{align}

\noindent with additional axion-independent terms that contribute to the vacuum energy, i.e. $\mathrm{Tr}(M_\Phi^\dagger M_\Phi)$ and $\mathrm{Tr}(M_\star^\dagger M_\star)$. If the terms in Eq.\eqref{eq:HP} are non-zero, the hierarchy problem is introduced in the theory and one has to understand how the axion can be naturally light.

However, the one-loop quadratic sensitivity is absent if
\begin{align}
    \mathrm{Tr}\big[M_\Phi^\dagger M_\star\big]=0,
    \label{eq:trcond}
\end{align}

\noindent while still having non-zero invariants at higher orders in $M_\nu$ to generate the potential. As usual, one should expect a symmetry reason why such term vanishes. To make this point clear, we can restore the flavor symmetries by assigning charges to the entries of the spurion matrices $M_\Phi$ and $M_\star$. The trace condition in Eq.\eqref{eq:trcond} is a selection rule imposed by $M_\Phi$ and $M_\star$ belonging to different charge sectors, which is the statement that they are orthogonal in flavor space. For example, a simple texture that realizes this selection rule is having $M_\Phi$ being off-diagonal and $M_\star$ diagonal in the interaction basis.

\subsubsection*{Flavor structure of the axion potential}

We now examine how the flavor structure of the neutrino mass matrix controls the radiatively generated axion potential. The orthogonality condition from Eq.~\eqref{eq:trcond} removes the quadratically divergent contribution to the axion potential, but the remaining radiative potential depends on the relative flavor alignment of the two neutrino mass terms. To study this dependence, we consider a rank-one axion-independent breaking term,
\begin{align}
    M_\star
    = \mu_\star
      \ket{n_\star}\bra{n_\star^\ast},
    \qquad
    \langle n_\star|n_\star\rangle = 1 .
    \label{eq:Mstar_rankone}
\end{align}
Here $\ket{n_\star}$ is a normalized flavor direction, whose
orientation will be specified below. The complex conjugation in
$\bra{n_\star^\ast}$ ensures that $M_\star$ is symmetric, as required
for a Majorana mass matrix. After fixing $M_{\nu,0}$ by Eq.\eqref{eq:Mnu_PMNS} with the PMNS parameters, $M_\Phi$ can be fully determined by
\begin{align}
    M_{\Phi} = \big( M_{\nu,0} - M_\star\big) e^{i a_0/f}.
    \label{eq:MPHI}
\end{align}

\noindent Thus, we can write the axion-dependent Majorana mass,
\begin{align}
    M_{\nu}(a) = \big( M_{\nu,0} - M_\star\big)
    e^{-i (a-a_0)/f} + M_\star.
\end{align}

\noindent We can take the trace in Eq.\eqref{eq:trcond} using
the rank-one texture and impose that it vanishes to get
\begin{align}
    \bra{n_\star} M_\Phi \ket{n_\star^\ast} = 0.
    \label{eq:nMPhin}
\end{align}

\noindent Substituting Eq.\eqref{eq:MPHI} into Eq.\eqref{eq:nMPhin},
we can fix $\mu_\star$,
\begin{align}
    \mu_\star
    = \bra{n_\star} M_{\nu,0}\ket{n_\star^\ast}
    = \sum_{i=1}^3 m_i
    \langle n_\star|\nu_i^\ast\rangle^2.
    \label{eq:mu_rankone}
\end{align}

\noindent Therefore, once the direction $\ket{n_\star}$ is specified, our choices for the rank-one texture and vanishing interference term \eqref{eq:trcond} are enough to fix $M_\Phi$ and $M_\star$ in terms of the parameters of the PMNS matrix and lightest neutrino mass.

Charged leptons can provide an additional source of flavor breaking that can modify the axion potential through weak interactions at higher loops. In the charged lepton mass basis, we define $D_\ell^2=\mathrm{diag}(m_e^2,m_\mu^2,m_\tau^2)$. Two-loop charged current diagrams involving a $W$ boson and charged leptons can generate contributions of the form
\begin{align}
    \Delta V_\ell^{(2)}(a)
    \sim
    \frac{g_2^2}{(16\pi^2)^2}
    \operatorname{Re}\!\left[
        e^{ia/f}
        \operatorname{Tr}
        \!\left(M_\Phi^\dagger M_\star D_\ell^2\right)
    \right].
    \label{eq:charged_lepton_potential}
\end{align}

\noindent The unequal charged lepton masses weight the flavor contributions differently, so Eq.~\eqref{eq:trcond} does not generically remove this invariant. The resulting contribution scales quadratically with the charged lepton masses and can dominate the quartic neutrino potential despite the additional loop suppression. For the rank-one examples that are aligned with a lepton flavor eigenstate,  $|n_\star\rangle=|\alpha\rangle$, with $\alpha=e,\mu,\tau$, $M_\star$ has only an $\alpha\alpha$ entry and Eq.~\eqref{eq:trcond} requires $(M_\Phi)_{\alpha\alpha}=0$. Consequently, 
\begin{align}
    \operatorname{Tr}
    \!\left(M_\Phi^\dagger M_\star D_\ell^2\right)
    =
    m_\alpha^2\mu_\star
    (M_\Phi)_{\alpha\alpha}^\ast
    =0.
\end{align}

\noindent In these examples, the contribution vanishes because $M_\Phi$ and $M_\star$ have disjoint support in flavor space,
\begin{align}
    (M_\Phi)_{ij}^\ast(M_\star)_{ij}=0 \quad \text{for every $i,j$}.
\end{align} 

\noindent Therefore, any charged-lepton insertions on either flavor index cannot restore the interference at quadratic order in the neutrino mass spurions. We therefore focus on flavor-aligned rank-one textures with this disjoint support. 

In Appendix~\ref{app:democratic}, we discuss a democratic flavor-breaking direction that requires additional protection against charged-lepton contributions. Its strong sensitivity to the current uncertainties in $\delta_{\rm CP}$ and $\theta_{23}$ illustrates how neutrino parameters can control the scale of the dark-energy potential.

With the quadratic terms controlled as above, the one-loop neutrino contribution starts at quartic order in the mass matrix.
Suppressing logarithmic factors in this leading-invariant estimate, the contribution with four mass insertions is
\begin{align}
    V_{\rm CW}(a)
    \simeq
    -\frac{g_\nu}{64\pi^2}
    \operatorname{Tr}\!\left[
        \big(M_\nu^\dagger(a)M_\nu(a)\big)^2
    \right].
    \label{eq:VCW_leading}
\end{align}

\noindent With $g_\nu=2$ for each Majorana mass eigenstate. The second harmonic vanishes for the rank-one texture satisfying Eq.~\eqref{eq:trcond}, leaving, up to axion-independent terms,
\begin{align}
    V_{\rm CW}(a)
    \simeq
    - \Lambda_\nu^4
    \cos\left(\frac{a-a_0}{f}+\delta_{\nu}\right),
\end{align}
where the amplitude and phase of the potential are
\begin{align}
    \Lambda_\nu^4 e^{i\delta_\nu}
    &\equiv \frac{e^{ia_0/f} }{8\pi^2} 
    \operatorname{Tr}\!\left[
        \big(M_\Phi^\dagger M_\Phi
            +M_\star^\dagger M_\star\big)
        M_\Phi^\dagger M_\star
    \right].
\end{align}

\noindent Defining the overlaps $q_i \equiv \langle n_\star|\nu_i^\ast\rangle$ and expanding the trace leads to
\begin{align}
    \Lambda_\nu^4
    &\equiv
    \left|
        \frac{\mu_\star}{8\pi^2}
        \sum_{i=1}^3 m_i
        \big[m_i q_i^\ast-\mu_\star^\ast q_i\big]^2
    \right|,
    \label{eq:VCW_rankone}
    \\
    \delta_\nu
    &\equiv
    \operatorname{Arg}\!\left[
        \mu_\star
        \sum_{i=1}^3 m_i
        \big[m_i q_i^\ast-\mu_\star^\ast q_i\big]^2
    \right].
\end{align}

\noindent The full logarithmic
Coleman--Weinberg potential, which can contain higher harmonics,
will be evaluated separately.

Before specifying a few flavor directions, we can identify some general properties of Eq.~\eqref{eq:VCW_rankone} by considering the hierarchical limits of normal and inverted ordering. These limits isolate the atmospheric scale as the leading contribution to the potential. 

For normal ordering, the limit $m1,m2$ going to zero at fixed $m_3 = m_{\rm atm}$ gives $\mu_\star = m_3 q_3^2$ and isolates the pure-atmospheric term
\begin{align}
    \Lambda_\nu^4 \big|_{\rm NO}
    \simeq
    \frac{m_{\rm atm}^4}{8\pi^2}
    |q_3|^4\big(1-|q_3|^2\big)^2 + \mathcal{O}\left(m_3^3 m_2\right).
    \label{eq:VCW_NO_atm}
\end{align}
This contribution is suppressed when the breaking direction is approximately orthogonal to the heaviest eigenstate, $q_3\simeq0$, or approximately aligned with it, $|q_3|\simeq1$. When this term is suppressed, mixed contributions of order $m_2m_3^3$ and terms involving the lightest mass can become important as we show in the examples below. 

For inverted ordering, taking $m_3\ll m_1\simeq m_2 =  m_{\rm atm}$ and neglecting the solar splitting at leading order, we obtain $\mu_\star\simeq m_{\rm atm}(q_1^2+q_2^2)$. The atmospheric contribution becomes
\begin{align}
    \Lambda_\nu^4 \big|_{\rm IO}
    \simeq
    \frac{m_{\rm atm}^4}{8\pi^2}
    \prod_{\sigma=\pm1}
    |q_1+\sigma i q_2|^2
    \left|1-|q_1+\sigma i q_2|^2\right| + \dots
    \label{eq:VCW_IO_atm}
\end{align}
Both heavy-state overlaps contribute at this order, and their relative phase can produce additional cancellations. Since $m_{\rm atm}^4/(8\pi^2)\gtrsim(17\,\mathrm{meV})^4$, reaching the dark-energy scale requires flavor-breaking directions that strongly suppress these leading atmospheric contributions.

A simple realization of this suppression is provided by the electron direction, $|n_\star\rangle=|e\rangle$, for which
\begin{align}
    M_\star \big|_e&=
    \mu_\star
    \begin{pmatrix}
        1 & 0 & 0 \\
        0 & 0 & 0 \\
        0 & 0 & 0
    \end{pmatrix},
    \qquad \mu_\star=(M_{\nu,0})_{ee}.
\end{align}

\noindent The corresponding axion-dependent neutrino mass matrix has a vanishing electron-electron entry, $(M_\Phi)_{ee}=0$, making the two mass matrices have disjoint support in flavor space, $(M_\Phi^*)^{ij}(M_\star)_{ij}=0$ for all $i,j$. 

In this case, $q_i=V_{ei}^\ast$ and $|q_3|^2=s_{13}^2$. For hierarchical normal ordering, a useful estimate retains the full $\mu_\star$ while keeping the $i=3$ term in the sum in Eq.~\eqref{eq:VCW_rankone},
\begin{align}
    \Lambda_\nu^{e,\mathrm{NO}}
    &\simeq
    \left[
      \frac{|\mu_\star|m_3}{8\pi^2}
      \left|m_3q_3^\ast-\mu_\star^\ast q_3\right|^2
    \right]^{1/4}.
    \label{eq:Lambda_e_NO_estimate}
\end{align}

\noindent The lighter states enter this estimate through the breaking coefficient,
\begin{align}
    \mu_\star
    &=
    c_{13}^{2}\left(
    m_1c_{12}^{2}
    +m_2s_{12}^{2}e^{-i\alpha_{21}}
    \right)
    +m_3s_{13}^{2}
    e^{i(2\delta_{\rm CP}-\alpha_{31})}.
\end{align}

\noindent For $m_1\simeq0$ and the central oscillation inputs, $m_2s_{12}^2c_{13}^2\simeq2.6\,\mathrm{meV}$, while $m_3s_{13}^2\simeq1.1\,\mathrm{meV}$. Thus, the solar term is larger in $\mu_\star$, even though the atmospheric term controls the remaining sum in Eq.~\eqref{eq:VCW_rankone}. For $|\mu_\star|\ll m_3$, this estimate scales as
\begin{align}
     \Lambda_\nu^{e,\mathrm{NO}}
     &\sim 2.4\,\mathrm{meV}
     \left[\frac{|\mu_\star|}{1\,\mathrm{meV}}\right]^{1/4}
     \left[\frac{m_3}{50\,\mathrm{meV}}\right]^{3/4}
     \left[\frac{s_{13}^2}{0.02}\right]^{1/4}.
     \label{eq:Lambda_e_NO_scaling}
\end{align}

\noindent Setting $m_1=m_2=0$ in Eq.~\eqref{eq:Lambda_e_NO_estimate} recovers Eq.~\eqref{eq:VCW_NO_atm} with $|q_3|^2=s_{13}^2$.

For example, choosing $\alpha_{21}=\pi$ and $\alpha_{31}=2\delta_{\rm CP}$ gives
\begin{align}
    \mu_\star
    &=
    m_1c_{12}^{2}c_{13}^{2}
    -m_2s_{12}^{2}c_{13}^{2}
    +m_3s_{13}^{2}.
\end{align}
The $m_2$ contribution then has opposite sign to the $m_1$ and $m_3$ contributions. For these phases and the central oscillation inputs, the cancellation occurs near $m_1\simeq2.35\,\mathrm{meV}$. Varying the lightest mass $m_1$, with $m_2$ and $m_3$ determined by the measured mass-squared splittings, can bring these contributions close to cancellation and further suppress the coefficient in Eq.~\eqref{eq:VCW_rankone}. Since $m_{\beta\beta}=|\mu_\star|$ for this direction, the same cancellation also suppresses the light-neutrino contribution to neutrinoless double beta decay.

However, a suppressed axion potential need not imply a small effective Majorana mass. For the muon direction, $|n_\star\rangle=|\mu\rangle$, we have
\begin{align}
    M_\star\big|_\mu
    &=
    \mu_\star
    \begin{pmatrix}
        0 & 0 & 0 \\
        0 & 1 & 0 \\
        0 & 0 & 0
    \end{pmatrix},
    \qquad
    \mu_\star=(M_{\nu,0})_{\mu\mu}.
\end{align}
The breaking coefficient and $m_{\beta\beta}=|(M_{\nu,0})_{ee}|$ now probe different entries of the neutrino mass matrix. To illustrate the suppression analytically, consider inverted ordering. The measured mass-squared splittings leave the lightest mass, $m_{\rm min}=m_3$, as a free parameter,
\begin{align}
    m_1^2 &= m_{\rm min}^2+|\Delta m_{31}^2|,
    \\
    m_2^2 &= m_{\rm min}^2+|\Delta m_{31}^2|
             +\Delta m_{21}^2.
\end{align}

\noindent Neglecting the solar splitting, we write $m_1\simeq m_2\simeq m_{\rm atm} \equiv\sqrt{m_{\rm min}^2+|\Delta m_{31}^2|}$. We neglect $s_{13}$ and choose the Majorana phases $\alpha_{21}=\pi$ and $\alpha_{31}=0$ to exhibit destructive interference in $(M_{\nu,0})_{\mu\mu}$. In this limit, the overlaps become
\begin{align}
    q_1=-s_{12}c_{23},\qquad
    q_2=-i c_{12}c_{23},\qquad
    q_3=s_{23}.
\end{align}
Consequently,
\begin{align}
    \mu_\star
    &\simeq
    -m_{\rm atm}c_{23}^{2}\cos(2\theta_{12})
    +m_{\rm min}s_{23}^{2},
    \\
    m_{\beta\beta}
    &\simeq m_{\rm atm}\cos(2\theta_{12}),
\end{align}
The breaking coefficient is therefore suppressed near
\begin{equation}
    m_{\rm min}
    \simeq
    m_{\rm atm}\frac{c_{23}^{2}}{s_{23}^{2}}
    \cos(2\theta_{12}),
\end{equation}
while $m_{\beta\beta}$ remains of order $20\,\mathrm{meV}$. Close to this cancellation, Eq.~\eqref{eq:VCW_rankone} gives
\begin{align}
    \Lambda_\nu^4\big|_{\mu,\rm IO}
    \simeq
    \frac{|\mu_\star|}{8\pi^2}\,
    m_{\rm min}s_{23}^{2}
    |\Delta m_{31}^2|.
\end{align}
Thus, the radiative potential can be suppressed without suppressing neutrinoless double beta decay.

These examples show that the neutrino masses set the underlying scale of the axion potential, but the PMNS mixing angles and CP phases also play a crucial role, controlling the flavor overlaps that suppress it to the dark-energy scale. Within the rank-one construction, specifying the protected flavor
direction and the complete present-day neutrino mass matrix fixes the neutrino-induced potential in the prescription used below. Its height therefore depends on the oscillation parameters, the lightest mass, and the Majorana phases.

\begin{figure}
    \centering
    \includegraphics[width=1\linewidth]{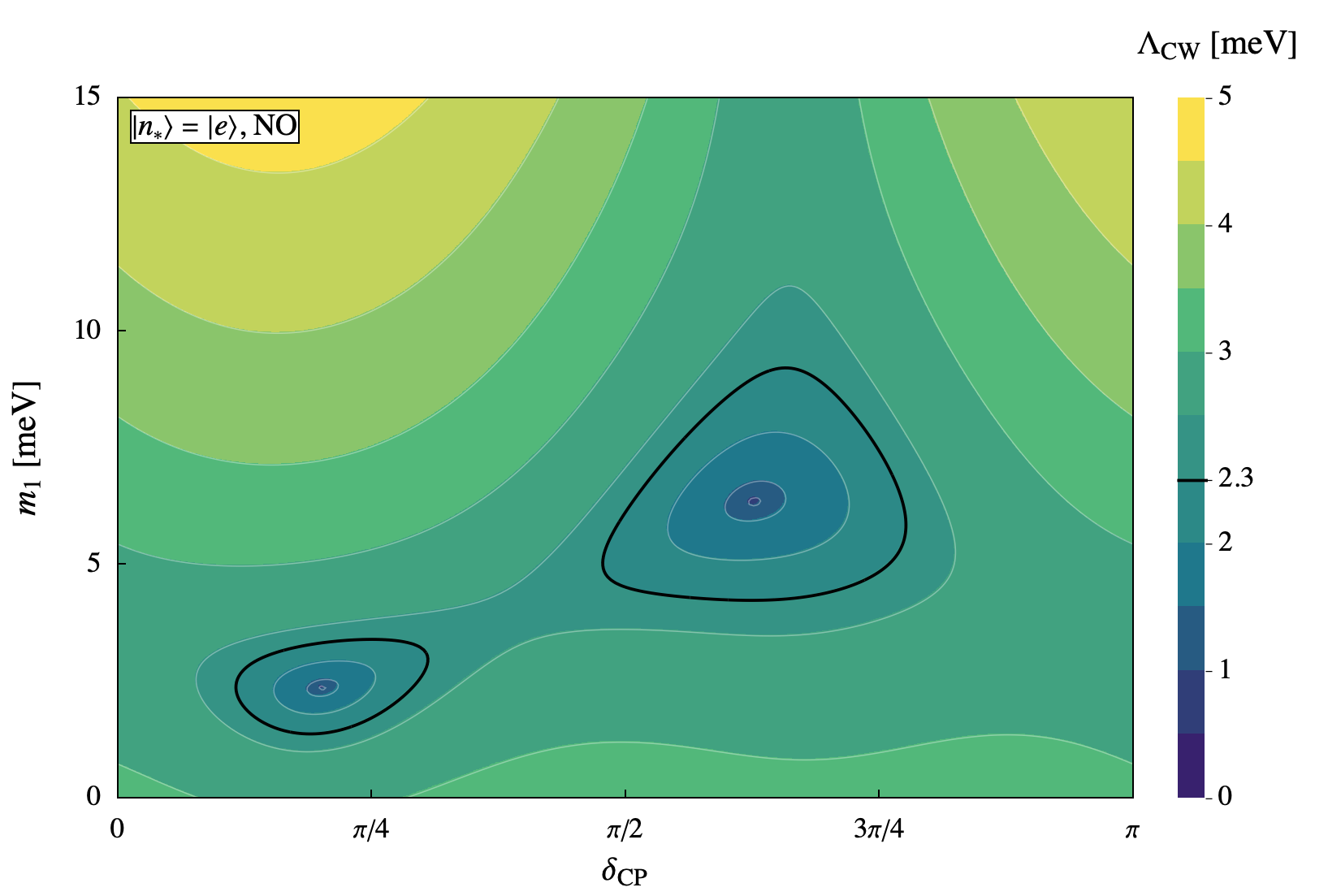}
    \caption{
        Dark energy potential scale $\Lambda_{\rm CW}=(V_{\max}-V_{\min})^{1/4}$, in meV, for the electron flavor breaking direction $|n_\star\rangle=|e\rangle$ and normal ordering, as a function of $\delta_{\rm CP}$ and $m_1$. The Majorana phases are fixed to $\alpha_{21}=175^\circ$ and $\alpha_{31}=60^\circ$, while the remaining oscillation parameters take their best-fit values. The black contour marks $\Lambda_{\rm CW}=2.3\,\mathrm{meV}$.}
    \label{fig:Lambda_e_NO}
\end{figure}

We now evaluate the full Coleman--Weinberg potential, retaining the dependence of the logarithms on the neutrino masses. For each value of $a$, we obtain the physical squared masses $m_{\nu_i}^2(a)$ by diagonalizing $M_\nu^\dagger(a)M_\nu(a)$. The resulting potential is
\begin{equation}
    V_{\rm CW}(a)=
    -\frac{g_\nu}{64\pi^2}
    \sum_{i=1}^3 m_{\nu_i}^4(a)
    \log\left(\frac{m_{\nu_i}^2(a)}{\bar\mu^2}\right),
    \label{eq:VCW_full}
\end{equation}
We absorb the usual finite $-3/2$ term into $\bar\mu$. We choose the renormalization scale to be the sum of squared neutrino masses independent of $a$, which is field independent due to the trace condition in Eq.~\eqref{eq:trcond}, 
\begin{equation}
    \bar\mu^2
    =\sum_{i=1}^3 m_{\nu_i}^2(a)
    =\operatorname{Tr}(M_\Phi^\dagger M_\Phi)
     +\operatorname{Tr}(M_\star^\dagger M_\star)
    =\sum_{i=1}^3 m_i^2 .
    \label{eq:sum_mass_squared}
\end{equation}

\noindent Since $\sum_i m_{\nu_i}^4(a)$ depends on the axion field, the potential retains some logarithmic sensitivity to the ultraviolet matching of the local operator
\begin{align}
    \operatorname{Re}\!\left[
    e^{ia/f}\operatorname{Tr}\!\left[
    \left(M_\Phi^\dagger M_\Phi+M_\star^\dagger M_\star\right)
    M_\Phi^\dagger M_\star
    \right]\right].
\end{align}
This term has the same flavor structure as Eq.~\eqref{eq:VCW_rankone} and vanishes as $\mu_\star\to0$, preserving the flavor suppression while modifying the potential height. Our numerical results assume that its renormalized coefficient vanishes at $\bar\mu^2=\sum_i m_i^2$. Imposing this condition at a higher scale $\Lambda_{\rm UV}$ instead adds a contribution of order $\Lambda_\nu^4\log(\Lambda_{\rm UV}^2/\bar\mu^2)$, which can move the dark energy scale solutions closer to the cancellation points of $\mu_\star$. With that, we subtract the minimum and characterize the height of the full potential by its total excursion,
\begin{alignat}{5}
    &V(a)
    &&\equiv V_{\rm CW}(a)-\min_a V_{\rm CW}(a),
    \\
    &\Lambda_{\rm CW}^4
    &&\equiv \max_a V_{\rm CW}(a)-\min_a V_{\rm CW}(a).
    \label{eq:LambdaCW}
\end{alignat}
The Majorana phases can affect both the position of the
minimum and the height of the potential, with a dependence
that varies with the breaking direction.

\begin{figure}
    \centering
    \includegraphics[width=1\linewidth]{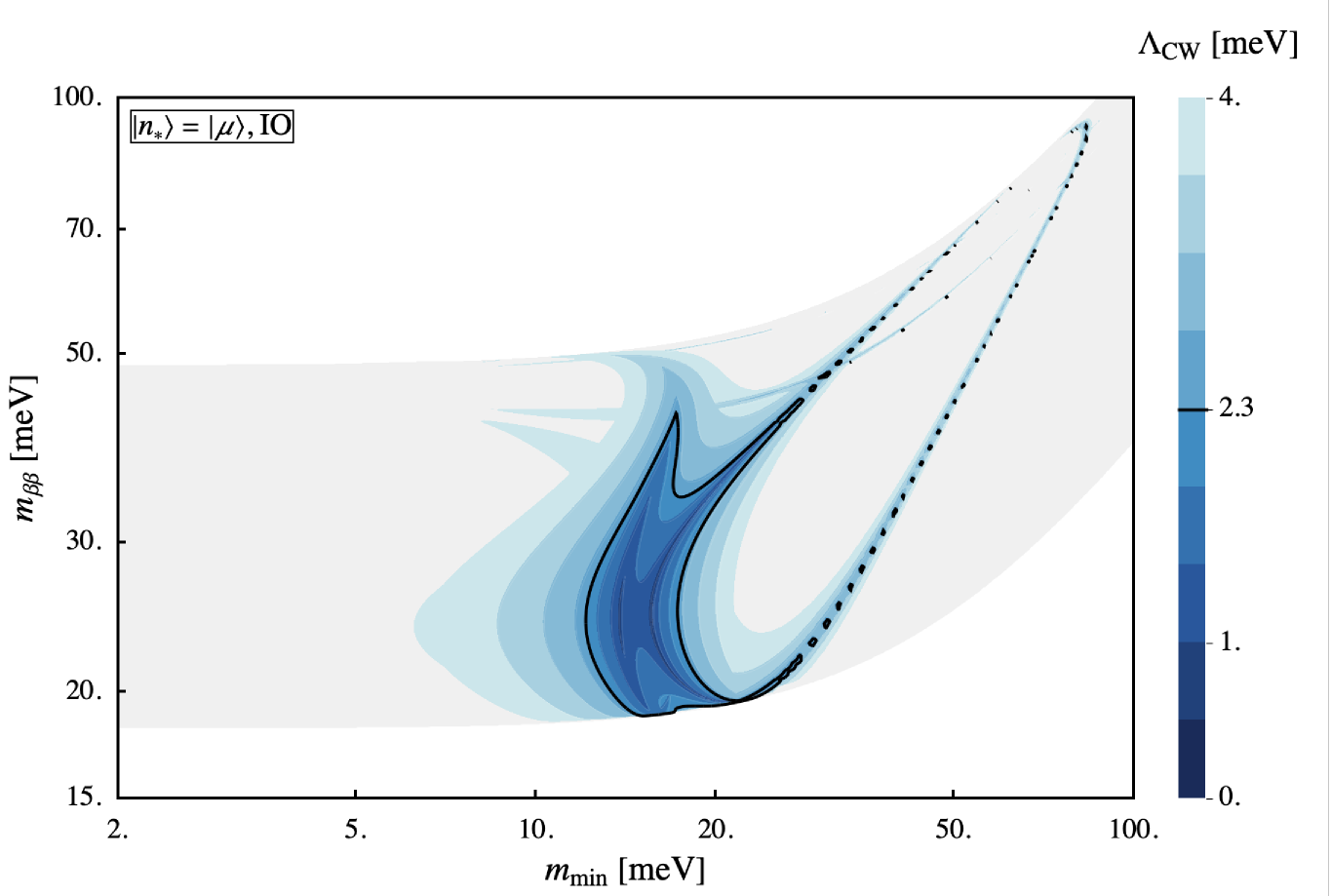}
    \caption{Dark energy potential scale in the $(m_{\min},m_{\beta\beta})$ plane for $|n_\star\rangle=|\mu\rangle$ and inverted ordering, with $m_{\min}=m_3$. Colors show the minimum of $\Lambda_{\rm CW}=(V_{\max}-V_{\min})^{1/4}$ over both Majorana phases at each point, with the oscillation parameters fixed to their best-fit values. The black contour marks a minimum of $2.3\,\mathrm{meV}$.}
    \label{fig:Lambda_mu_IO}
\end{figure}

The examples illustrate different connections between the radiative scale and neutrino observables. For the electron direction, $|n_\star\rangle=|e\rangle$, Fig.~\ref{fig:Lambda_e_NO} shows the full $\Lambda_{\rm CW}$ in the $(\delta_{\rm CP},m_1)$ plane for normal ordering. We choose $(\alpha_{21},\alpha_{31})=(175^\circ,60^\circ)$, close to the destructive-interference configuration $(\pi,2\delta_{\rm CP})$ at the NuFIT best-fit value $\delta_{\rm CP}=212^\circ$ for illustrative purposes. The $m_2$ contribution then nearly opposes the $m_1$ and $m_3$ contributions, allowing partial cancellation as the lightest mass is varied. We also take the remaining oscillation parameters at their best-fit values. Changing $m_1$ modifies the relative weights of the neutrino contributions, while $\delta_{\rm CP}$ changes their interference through the combination $\alpha_{31}-2\delta_{\rm CP}$. The black contours, corresponding to $\Lambda_{\rm CW}=2.3\,\mathrm{meV}$, therefore select correlations between the lightest mass and the Dirac phase for this choice of Majorana phases.

\noindent A different connection is illustrated by the
muon direction, $|n_\star\rangle=|\mu\rangle$, in
inverted ordering. Fig.~\ref{fig:Lambda_mu_IO} presents
the result in the $(m_{\min},m_{\beta\beta})$ plane,
where $m_{\min}=m_3$ and
\begin{equation}
    m_{\beta\beta}
    =\left|\sum_{i=1}^3 m_i V_{ei}^2\right|
    =\left|(M_{\nu,0})_{ee}\right|.
\end{equation}
At each point, the colours show the minimum of
$\Lambda_{\rm CW}$ over both Majorana phases, with the
oscillation parameters fixed to their best-fit values.
The black contour marks a minimum of
$2.3\,\mathrm{meV}$, while grey denotes the allowed
region where this minimum exceeds $4\,\mathrm{meV}$. In this case, the breaking coefficient is
$\mu_\star=(M_{\nu,0})_{\mu\mu}$, whereas neutrinoless
double beta decay probes $(M_{\nu,0})_{ee}$.
Consequently, suppressing the radiative potential does
not require a small effective Majorana mass.
The figure exhibits dark-energy-scale potentials
together with $m_{\beta\beta}$ of tens of meV.
For the standard light-Majorana-neutrino exchange
mechanism, this lies in the region targeted by
next-generation neutrinoless double beta decay
searches~\cite{Adams:2022jwx}.

An analogous contribution to the ones induced by the Weinberg operators can arise from the bare $B+L$ violating operator
\begin{align}
    \mathcal{L}_{BLV} = \frac{C_B}{\Lambda_B^2} QQQL + h.c.
\end{align}

\noindent This operator then interferes with the $QQQL\Phi$ term in Eq.\eqref{eq:lag} and contributes to an effective potential for the axion as in the neutrino case. However, the scale of such an operator is tightly constrained from the bounds of proton decay, which leads to $\Lambda_B/||C_B||^{1/2} \gtrsim 10^{15} \GeV$  \cite{ParticleDataGroup:2024cfk}. The induced proton--lepton mixing gives a negligible low-energy contribution to the axion potential. Ultraviolet-sensitive interference between the two baryon-violating operators must also be suppressed by suitable flavor selection rules, independently of the neutrino condition in Eq.~\eqref{eq:trcond}.

Finally, the axion mass is determined by the
curvature of the full potential,
\begin{equation}
    m_a^2=
    \left.
    \frac{\partial^2 V_{\rm CW}}{\partial a^2}
    \right|_{a=a_{\min}} .
\end{equation}
In the single-cosine limit,
$m_a^2=\Lambda_{\rm CW}^4/f^2$.
A potential at the meV scale and a decay constant near
the Planck scale can therefore give $m_a\sim H_0$,
allowing the field to evolve at late times.
In Sect.~\ref{sec:III}, we discuss the preferred decay
constant and the resulting equation of state.

\subsection{Ultraviolet model requirements}
\label{sec:IID}

The effective theory above requires two inequivalent sources of the Weinberg operator with different axion dependence and a protected flavor structure. A microscopic completion must explain their origin and alignment while suppressing additional contributions to the axion potential. Here we outline these requirements without committing to a particular UV model.

A possible origin of the flavor structure is to decompose the lepton doublets into a flavor singlet $L_\star$, associated with the direction $|n_\star\rangle$, and a doublet $L_F^i$ spanning the orthogonal flavor plane. Promoting $\Phi$ to a scalar multiplet $\Phi^i$ in the same real doublet representation allows it to contract with doublet components of the Weinberg operator, including $(L_\star H)(L_F^iH)$.  The axion is the common phase of this multiplet, $\Phi^i\propto e^{-ia/f}$. In contrast, the singlet operator $(L_\star H)^2$ cannot couple linearly to $\Phi^i$ without an additional flavor spurion. Schematically, the two sources can therefore arise from
\begin{align}
    \mathcal L_\nu \supset
    \frac{\Phi^i}{f^2}(LHLH)_{\mathbf{2},i}
    +\frac{c_\star}{\Lambda_\star}(L_\star H)^2
    +{\rm h.c.},
\end{align}
where the first term denotes the allowed doublet contractions with their coefficients. Assigning lepton number $-2$ to $\Phi^i$ makes this term lepton-number conserving, whereas the axion-independent singlet term explicitly violates lepton number by two units.

This selection rule separates the singlet breaking term from the $\Phi$-dependent operators and gives $\operatorname{Tr}(M_\Phi^\dagger M_\star)=0$ automatically. Its essential ingredient is the distinction between a flavor direction and its orthogonal plane, rather than a particular choice of flavor group. The rank-one form of $M_\star$ adopted above is a simplifying assumption to isolate the effects of the single flavor-breaking direction.  More general textures can arise from additional axion-independent Weinberg operators, including singlet contractions within the doublet plane. These possibilities extend the framework beyond the examples considered here, provided that they reproduce the observed neutrino masses and mixing and preserve the required radiative protection of the axion potential.

The operator $QQQL\Phi$ is not required for the neutrino-driven dark-energy mechanism and plays a different role from the Weinberg operators. In the minimal construction, where only the Weinberg operator is dressed by the axion, $LHLH\Phi$, the weak axion is a Majoron associated with a purely leptonic symmetry. This is sufficient to generate the neutrino-induced potential at the dark-energy scale, and the anomalous $B+L$ interpretation is not essential. The role of $QQQL\Phi$ is instead to provide a possible $B+L$ embedding. Although this operator can saturate the fermionic zero modes of the electroweak ’t Hooft vertex, the corresponding instanton-induced potential is far below the dark-energy scale and is negligible for the mechanism studied here. Its main purpose is to fix the axion charge direction: the presence of both $LHLH\Phi$ and $QQQL\Phi$ selects the anomalous $B+L$ charge for $\Phi$, distinguishing the weak axion from a $B-L$ or purely leptonic axion. Possible phenomenological implications for baryon-number dynamics, such as spontaneous-baryogenesis-like effects, are left for future work.

A separate, more speculative possibility is to identify the $B+L$ weak axion with the $\eta_W'$ state proposed in Refs.~\cite{Dvali:2024zpc,Dvali:2025pcx,Kobakhidze:2026ymc}. In that framework, the additional pseudoscalar screens the topological susceptibility associated with the weak vacuum angle. This motivation relies in part on conjectural assumptions about de Sitter vacua in quantum gravity, and it is not required for the construction presented here. Independently of this interpretation, the weak axion effective theory is useful because it makes explicit the physical combination of $\theta_W$ and the phases of the $B+L$-violating spurions, thereby clarifying the structure of the instanton-induced potential.

For a potential of height $\rho_{\rm DE}$, the characteristic curvature is of order $\rho_{\rm DE}/f^2$. Keeping the axion slowly evolving today, without choosing an initial position very close to a maximum, therefore favors $f$ of order the Planck scale. However, this does not force the seesaw scale to be Planckian. If $f$ is set by the expectation value of $\Phi$, right-handed neutrinos coupled to $\Phi^\ast$ can acquire masses
\begin{align}
    M_N\sim\lambda f,
    \qquad
    m_\nu\sim\frac{y_\nu^2v^2}{2M_N},
\end{align}
where $\lambda$ denotes their coupling to $\Phi^\ast$, $y_\nu$ is a typical Dirac Yukawa coupling, and $v=246\,\mathrm{GeV}$. For $m_\nu\sim0.05\,\mathrm{eV}$ and $y_\nu\sim1$, this gives $M_N\sim6\times10^{14}\,\mathrm{GeV}$. The usual seesaw scale is thus compatible with $f\sim M_{\rm pl}$ for $\lambda\sim10^{-4}$--$10^{-3}$. In the normalization of Eq.~\eqref{eq:lag}, $C_\nu$ is correspondingly large, reflecting $M_N\ll f$ rather than a large fundamental Yukawa coupling.

The large $f$ entering the neutrino mass matrix could also be an effective scale, larger than the individual scalar expectation values. In clockwork models, several axions mix so that a light combination couples to a chosen sector with an enhanced effective decay constant \cite{Kaplan:2015fuy}. This could allow a near-Planckian $f$ while keeping the underlying symmetry-breaking scales lower. The neutrino operators would need to couple to this light combination with the enhanced period, while the heavy-neutrino masses would depend on the individual expectation values and their couplings. Other interactions must also be controlled, since they can generate shorter-period contributions that spoil the slow evolution. 

As in other axion models, the ultraviolet completion must suppress additional sources of explicit shift-symmetry breaking. Operators schematically of the form
\begin{align}
    \Delta V
    \sim
    \frac{c_n}{M_{\rm pl}^{n-4}}\Phi^n
    + {\rm h.c.}
\end{align}

\noindent generate an axion potential after $\Phi$ condenses whose amplitude is independent of the neutrino masses. Keeping these contributions sufficiently small is an instance of the axion quality problem familiar from QCD axion models. This requirement is particularly restrictive when the underlying scalar expectation values approach $M_{\rm pl}$, since suppression by powers of the Planck mass alone is then insufficient. Axions arising from gauge fields in extra dimensions and higher-form symmetries offer possible protection against explicit breaking \cite{Choi:2003wr,Craig:2024dnl}. Additional ultraviolet contributions, including instanton effects, must remain small over the required field range. After subtracting an axion-independent constant, we require
\begin{align}
    \left|\Delta V_{\rm UV}\right|
    &\lesssim \rho_{\rm DE},
    &
    \left|\Delta V_{\rm UV}''\right|
    &\lesssim H_0^2,
    \label{eq:quality_conditions}
\end{align}

\noindent where primes denote derivatives with respect to $a$. These bounds define the required axion quality. A microscopic realization satisfying them together with the neutrino couplings and flavor structure is left for future work.

\section{Cosmological evolution}
\label{sec:III}
The zero-temperature Coleman-Weinberg potential derived above fixes the weak axion mass and potential height in terms of the neutrino sector. The remaining cosmological inputs are the initial axion displacement, axion decay constant scale and the properties of the neutrino background. Since the neutrino masses depend on the axion field, the homogeneous axion evolves in a neutrino medium whose properties, in principle, change with redshift. These can lead to finite-density corrections to the ordinary quintessence-like dynamics of the scalar field. However, as we discuss next, the condition $\sum_i m_i^2(a)={\rm constant}$ removes the leading relativistic finite-density force.

Since $f\sim M_{\rm pl}$, the $B+L$ order parameter should be already present during inflation. The axion is then a light spectator field and inflation homogenizes its value over the observable universe, selecting a misaligned initial value $a_i$. After reheating, the SM plasma thermalizes, but the weak axion does not. Its largest thermal production channels come from the anomalous coupling to electroweak gauge bosons, which give the parametric rate
\begin{align}
    \Gamma_{aW} \sim \frac{\alpha_w^3 T^3}{f^2}.
\end{align}

\noindent During radiation domination,
\begin{align}
    \frac{\Gamma_{aW}}{H} \sim \frac{\alpha_w^3}{1.66 \sqrt{g_*}}\frac{T M_{\rm pl}}{f^2} \ll 1
\end{align}

\noindent even when evaluated at the largest sub-Planckian temperatures reached after inflation. Direct couplings to neutrinos are still small, since $g_{a\nu\nu} \sim m_\nu/f$. The weak axion never reaches thermal equilibrium and does not give appreciable contributions to $\Delta N_{\rm eff}$.

The background evolution of the axion is governed by the homogeneous Klein-Gordon equation in an expanding Universe,
\begin{align}
&\ddot a+3H\dot a + V_{\rm eff}'(a,T_\nu) = 0.
\end{align}

\noindent The effective potential,
\begin{align}
    &V_{\rm eff}(a,T_\nu) \equiv V_{\rm CW}(a)+V_{\rm C\nu B}(a,T_\nu),
\end{align}

\noindent contains the cosmic neutrino background contribution through the neutrino energy density that evolves with the axion-dependent masses. Equivalently, its contribution to the axion force can be written in terms of the trace of the neutrino energy-momentum tensor,
\begin{align}
&V'_{\rm C\nu B}(a,T_\nu) = \sum_{i=1}^3 Q_i(a)\left(\rho_{\nu_i}-3P_{\nu_i}\right),
\qquad
\\
&Q_i(a)\equiv \partial_a \log m_{\nu_i}(a).
\end{align}

\noindent For a relic neutrino distribution, the trace for each mass eigenstate is
\begin{align}
\rho_{\nu_i}-3P_{\nu_i} = \frac{g_\nu T_\nu^4}{2\pi^2}
y_i^2
\int_0^\infty dq
\frac{q^2}{\sqrt{q^2+y_i^2}}
\frac{1}{e^q+1},
\end{align}

\noindent where $y_i\equiv m_{\nu_i}(a)/T_\nu$. 

In the high-temperature regime (HT), $m_{\nu_i}(a)\ll T_\nu$, the trace for each eigenstate becomes,
\begin{align}
\rho_{\nu_i}-3P_{\nu_i}
\stackrel{\text{HT}}{\simeq}
\frac{m_{\nu_i}^2(a)T_\nu^2}{12}
+
\frac{m_{\nu_i}^4(a)}{8\pi^2}
\left[
\log\left(\frac{m_{\nu_i}^2(a)}{T_\nu^2}\right)
+c_{\rm FD}
\right]
\end{align}

\noindent with $c_{\rm FD}=1+2\gamma_E-2\log\pi$. The zero-temperature Coleman-Weinberg contribution may be written as
\begin{align}
V'_{\rm CW}(a) = -\frac{g_\nu}{16\pi^2}
\sum_i Q_i(a)m_{\nu_i}^4(a)
\left[
\log\left(\frac{m_{\nu_i}^2(a)}{\mu^2}\right)
- 1
\right].
\end{align}

\noindent Combining the Coleman-Weinberg force with the high-temperature finite-density source gives
\begin{align}
V'_{\rm eff} \stackrel{\rm HT}{\simeq} \frac{g_\nu T_\nu^2}{48}
\partial_a \sum_i m_{\nu_i}^2(a)
+
\mathcal{O}\big(m_{\nu_i}^4(a)\big)
\end{align}

\noindent The leading term is the potentially large thermal force of order $T_\nu^2 m_\nu^2$. However, in our construction this term vanishes because the sum of the squared masses is independent of the axion field, i.e. Eq.~\eqref{eq:sum_mass_squared}. Therefore the leading high-temperature finite-density force cancels. The remaining contribution is of order $m_\nu^4$,
\begin{align}
V'_{\rm eff} \stackrel{\rm HT}{\simeq} \frac{g_\nu}{16\pi^2}
\left[
\log\left(\frac{\mu^2}{T_\nu^2}\right)
+c_{\rm eff}
\right]
\sum_i Q_i(a)m_{\nu_i}^4(a) .
\end{align}

\noindent Notice that the field-dependent logarithms $\log m_{\nu_i}^2(a)$ cancel between the finite-density and zero-temperature Coleman-Weinberg pieces. Thus, after the cancellation of the $T_\nu^2 m_\nu^2$ term, the first surviving axion-dependent correction is parametrically of order $m_\nu^4$, with a temperature-dependent coefficient controlled by $\log(\mu^2/T_\nu^2)$.

In the low-temperature regime (LT), $m_{\nu_i}(a)\gg T_\nu$, the pressure is negligible and the trace approaches the rest-mass density. For each eigenstate,
\begin{align}
\rho_{\nu_i}-3P_{\nu_i}
\stackrel{\rm LT}{\simeq}
m_{\nu_i}(a)n_{\nu_i}.
\end{align}

\noindent where the relic number density is
\begin{align}
n_{\nu_i} = \frac{3g_\nu\zeta(3)}{4\pi^2}T_\nu^3 .
\end{align}

\noindent The leading non-relativistic finite-density source is therefore
\begin{align}
V'_{\rm C\nu B}(a,T\nu)
\stackrel{\rm LT}{\simeq}
\sum_i Q_i(a)m_{\nu_i}(a)n_{\nu_i}
\end{align}

\noindent This term is proportional to the axion dependence of the neutrino rest masses and redshifts as the relic number density, $n_\nu\propto T_\nu^3\propto (1+z)^3$. For an order-of-magnitude estimate, suppose that the same neutrino eigenstate dominates both the finite-density and Coleman--Weinberg forces. Denoting its mass by $m_{\nu_h}$ and neglecting order-one logarithmic factors, we obtain
\begin{align}
    \frac{|V'_{\rm C\nu B}(a,z)|}{|V'_{\rm CW}(a)|}
    \sim
    12\zeta(3)
    \left(\frac{T_{\nu,0}}{m_{\nu_h}(a)}\right)^3
    (1+z)^3 .
    \label{eq:CnuB_CW_force_ratio}
\end{align}
For $m_{\nu_h}\simeq50\,\mathrm{meV}$, this gives approximately
$6\times10^{-7}(1+z)^3$. This estimate applies only after the relevant neutrino eigenstate has become non-relativistic. It shows that the medium force is strongly suppressed today because $T_{\nu,0}\ll m_{\nu_h}$. Evolving backward within the non-relativistic regime, the ratio grows as $(1+z)^3$, reflecting the increasing relic number density. The effect is therefore largest near the relativistic-to-non-relativistic transition. 

From Eq.~\eqref{eq:CnuB_CW_force_ratio}, the instantaneous force from cosmic neutrinos can compete with the Coleman-Weinberg force at high redshifts like $z \sim 100$. However, the axion does not adiabatically track the instantaneous minimum since at high redshift its evolution is still dominated by Hubble friction. Concretely, the effective mass of the axion at high $z$ is always smaller than the Hubble parameter,
\begin{align}
    \frac{m_{a, \rm eff}^2}{H^2}\sim 3 \big[ \Omega_{\rm DE}(z)+ \Omega_\nu(z)\big] \left( \frac{M_{\rm pl}}{f}\right)^2 \ll 1,
\end{align}

\noindent where $m_{a, \rm eff}^2 = V''_{\rm CW} + V''_{C\nu B}$. Therefore, the field begins to roll only at late times, when $m_{\rm eff}^2/ H_0^2 \sim 3 \Omega_{DE,0}(M_{\rm pl}/f)^2$. Thus, the finite-density contribution from the cosmic neutrino background does not drive the field in any stage, and the background evolution of the axion is close to that of standard thawing quintessence. The field is frozen by Hubble friction in the early Universe and evolves appreciably only near the present epoch.

This distinction is important because the usual MaVaNs instability arises precisely in the opposite limit, where the scalar rapidly adjusts to the neutrino-density-dependent minimum and mediates an additional attractive force among non-relativistic neutrinos. This behavior explains why the usual adiabatic instability of the mass-varying neutrino models is absent \cite{Afshordi:2005ym}. The zero-temperature Coleman-Weinberg potential and finite density C$\nu$B contribution are not independent forces, both are generated by the same axion dependence to the neutrino masses. The C$\nu$B contribution can only reweight this dependence by the relic neutrino abundance, set by $T_\nu$, and by the allowed neutrino masses. For the physical values, the neutrino backreaction is never large enough to pull the axion away from its inflationary misalignment angle into a neutrino controlled tracking solution. The weak axion therefore avoids the adiabatic MaVaNs regime and the associated neutrino-clustering instabilities.

\begin{figure}
    \flushleft
    \includegraphics[width=0.91\linewidth]{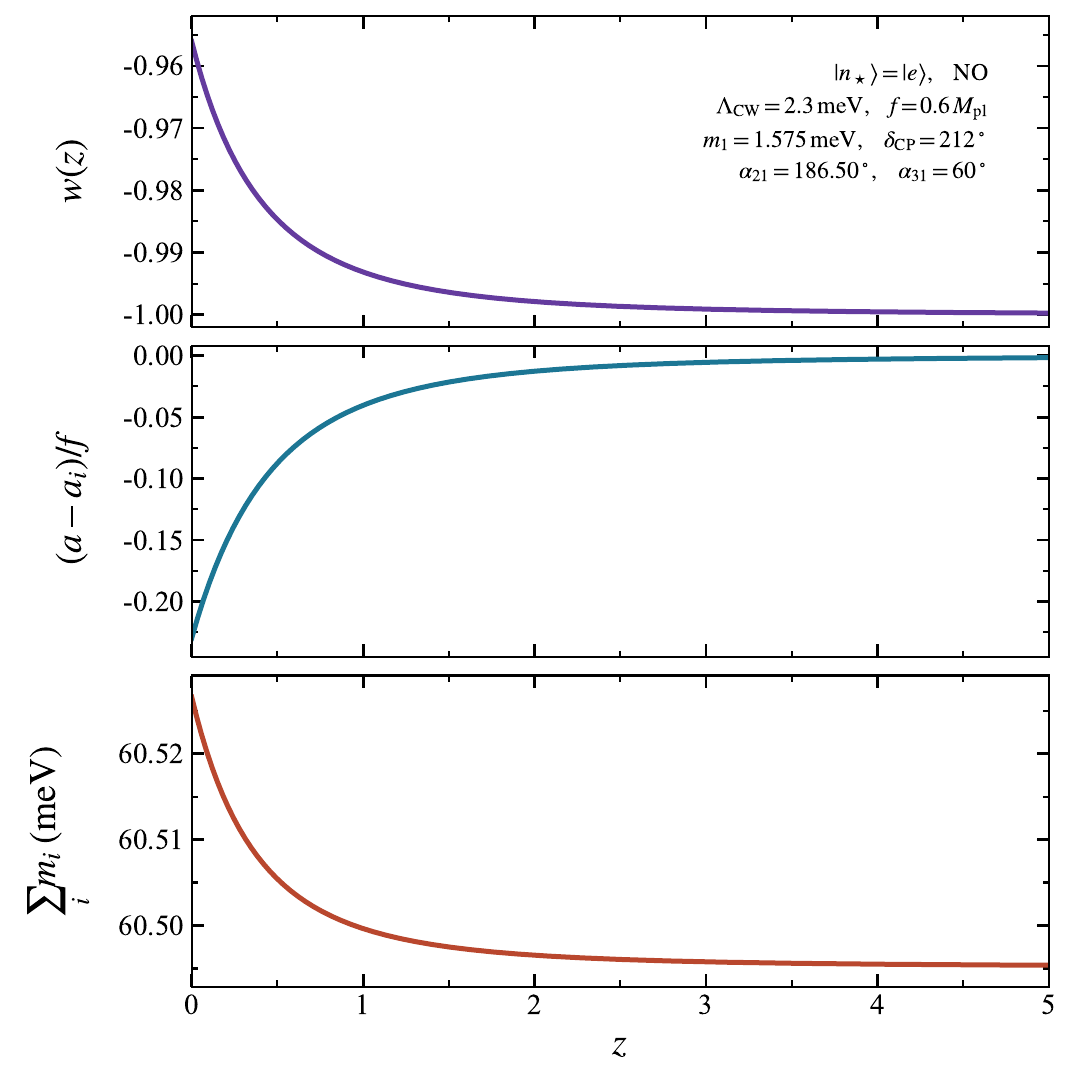}
    \caption{ Background evolution for the electron direction, $|n_\star\rangle=|e\rangle$, in normal ordering. The panels show the dark-energy equation of state $w(z)$, the axion displacement $(a-a_i)/f$, and the neutrino mass sum $\sum_i m_i$. The background uses a cosine potential with $\Lambda_{\rm CW}=2.3\,\mathrm{meV}$ and $f=0.6M_{\rm pl}$. The neutrino masses are evaluated along the resulting trajectory, neglecting their feedback because the finite-density force is much smaller than the Coleman--Weinberg force for this benchmark.
    }
    \label{fig:EoS}
\end{figure}

For the numerical illustration, we compute the thawing-quintessence background with \texttt{CLASS}, using a cosine approximation to the Coleman--Weinberg potential and keeping the neutrino masses fixed at their present-day values. The dark-energy equation of state is
\begin{align}
    w(z)=\frac{p_\phi(z)}{\rho_\phi(z)}
    =\frac{\frac{1}{2}\dot a^2-V_{\rm CW}(a)}
          {\frac{1}{2}\dot a^2+V_{\rm CW}(a)}.
\end{align}

Figure~\ref{fig:EoS} shows the evolution for the electron direction, $|n_\star\rangle=|e\rangle$, in normal ordering, with $\Lambda_{\rm CW}=2.3\,\mathrm{meV}$ and $f=0.6M_{\rm pl}$. The neutrino parameters shown in the inset reproduce the height and phase of the cosine approximation. Evaluating the field-dependent neutrino masses along this trajectory, we find that $\sum_i m_i$ changes by approximately $0.03\%$ between $z=5$ and today. The finite-density force is also much smaller than the Coleman--Weinberg force, supporting the neglect of neutrino feedback for this example.

\begin{figure}
    \centering
    \includegraphics[width=1\linewidth]{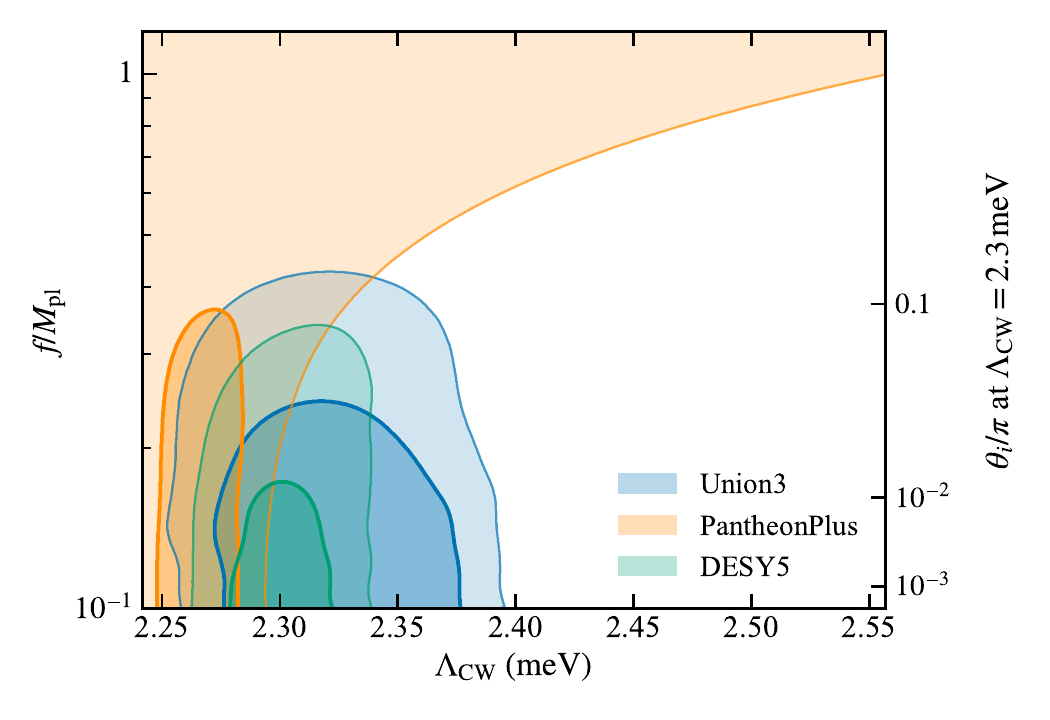}
    \caption{
        Confidence regions adapted from the two-parameter algebraic thawing analysis of Ref.~\cite{DESI:2025fii}.
        The shaded contours show the DESI constraints transformed from the parameters of the algebraic thawing equation of state \eqref{eq:algebraicEoS} to the $(\Lambda_{\rm CW}, f)$ plane.
        The right axis indicates the corresponding initial phase tuning, $\theta_i/\pi$, evaluated at
        $\Lambda_{\rm CW}=2.3\,{\rm meV}$.
    }
    \label{fig:DESI}
\end{figure}

To compare this benchmark with the DESI dynamical-dark-energy results, we recast the DESI extended analysis of algebraic thawing quintessence \cite{DESI:2025fii} into the weak-axion $(f,\Lambda_{\rm CW})$ parameter plane. The DESI analysis uses the two-parameter thawing form
\begin{align}
    1+w(a)
    =
    (1+w_0)\,a^p
    \left(
    \frac{1+b}{1+b\,a^{-3}}
    \right)^{1-p/3},
    \quad b=0.5 ,
    \label{eq:algebraicEoS}
\end{align}

\noindent which enforces $w(a)\geq -1$. For each point in the $(\Lambda_{\rm CW},f)$ plane, we solve the axion background evolution, tune the initial phase so that the scalar accounts for the observed dark energy density today, and fit the resulting $w(a)$ to the algebraic thawing form above. We then use this map to recast the DESI posterior regions in the $(w_0,p)$ plane into the $(\Lambda_{\rm CW},f)$ plane. The resulting recast regions are shown in Fig.~\ref{fig:DESI}.

In the neutrino construction, $\Lambda_{\rm CW}$ depends on the neutrino masses, PMNS mixing angles and CP phases, with their relative importance determined by the flavor-breaking direction. The present uncertainties in these parameters allow a range of potential heights and hence different cosmological evolutions. Improved neutrino measurements can therefore sharpen the connection between the predicted potential and thawing-quintessence constraints. We will explore this connection in a forthcoming detailed study, including the dependence on the flavor-breaking direction and the neutrino parameters.

\section{Conclusions}
\label{sec:V}

In this paper we explored the model-building requirements and phenomenology associated with the weak axion as a dark energy candidate. Dynamical dark energy is an attractive possibility because it can lead to observable departures from a cosmological constant, but it is difficult to realize naturally. The potential must have height $(2.3~{\rm meV})^4$, curvature of order $H_0^2$, and field excursions near the Planck scale. The weak axion addresses this hierarchy by tying its shift-symmetry breaking to the small sources of explicit $B+L$ violation.

We revisited the weak instanton-induced potential calculation and found that it is far below the observed dark energy density. Electroweak symmetry breaking screens large instantons through the Higgs VEV, and a nonzero contribution requires saturating the twelve fermionic zero modes of the electroweak 't Hooft operator. In the EFT considered here, this requires insertions of $QQQL$-type operators and leads to an extremely suppressed amplitude. Weak instantons therefore identify the anomalous $B+L$ direction, but they do not set the dark energy scale.

The dominant potential instead comes from the neutrino sector. The leading low-dimensional source of $B+L$ violation is the Weinberg operator, and two inequivalent $LHLH$ sources, one axion-dependent and one axion-independent, generate an axion-dependent Majorana mass matrix. If the two Weinberg spurions belong to different flavor-charge sectors, $\mathrm{Tr}(M_\Phi^\dagger M_\star)=0$, the dangerous quadratic contribution to the axion potential is absent. The leading radiatively stable Coleman--Weinberg potential then appears only at quartic order in Majorana mass insertions. The same condition also removes the leading high-temperature finite-density force proportional to $T_\nu^2m_\nu^2$.

Several flavor-breaking directions can suppress the neutrino-induced Coleman--Weinberg potential to the $\mathrm{meV}$ scale. The electron direction in normal ordering and the muon direction in inverted ordering show how PMNS mixing angles and interference between neutrino contributions control this suppression. The dependence on the lightest neutrino mass and CP phases differs between textures, with the Majorana phases affecting both the potential height and the position of its minimum.

The assumed flavor structure also determines the radiative stability of these results. Although Eq.~\eqref{eq:trcond} removes the unweighted quadratic invariant, unequal charged lepton masses can spoil this cancellation in higher-loop contributions. The electron- and muon-aligned textures have disjoint support for $M_\Phi$ and $M_\star$ in the charged lepton mass basis, eliminating this invariant. Appendix~\ref{app:democratic} discusses the democratic
direction and the additional protection that it requires.

For textures with the required radiative protection, the neutrino-induced potential connects dark energy
to several goals of neutrino physics.The electron direction in normal ordering can favor a small $m_{\beta\beta}$, whereas the muon direction in inverted ordering allows a suppressed potential with $m_{\beta\beta}$ in the tens-of-$\mathrm{meV}$ range. Proposed searches such as LEGEND-1000, nEXO and CUPID aim to probe the inverted-ordering region \cite{LEGEND:2021bnm,nEXO:2021ujk,CUPID:2025faj}, providing complementary tests of these possibilities.
Together, oscillation and neutrinoless double beta decay measurements can constrain the neutrino parameters that yield the dark-energy scale for each flavor structure. We plan to investigate the cosmological evolution and observational consequences of the protected flavor textures in forthcoming work.

Cosmologically, the model behaves as ordinary thawing quintessence for $f$ close to $M_{\rm pl}$. The weak axion is not thermalized in the early Universe, remains frozen by Hubble friction until late times, and is never driven by the cosmic neutrino background. After neutrinos become non-relativistic, the finite-density force redshifts with the relic abundance and remains too small to produce a neutrino-controlled minimum. The model therefore avoids the adiabatic regime of mass-varying neutrino scenarios and the associated nonlinear neutrino-clustering instability.

Several theoretical questions remain. We have not addressed the cosmological constant problem, and a complete UV construction must explain why $f$ can be near  $M_{\rm pl}$ while preserving a high-quality axion shift symmetry. It should also provide the flavor selection rules that remove the quadratic sensitivity of the potential and explain the origin of the flavor breaking spurion. Nevertheless, the weak axion gives a concrete framework in which the hierarchy required for quintessence is generated from known small sources of Standard Model symmetry breaking. This makes it a useful target for further studies of dynamical dark energy, neutrino masses, anomalous symmetries, and possible gravitational sources of flavor structure.

\vspace{0.5cm}
\acknowledgments
The authors thank Subhojit Roy, Hengameh Bagherian, Junwu Huang and Asimina Arvanitaki for useful discussions. Research at Perimeter Institute is supported by the Government of Canada through the Department of Innovation, Science and Economic Development Canada and by the Province of Ontario through the Ministry of Research, Innovation and Science. The work of C.W. at the University of Chicago has been supported by the DOE grant DE-SC0013642. C.W.'s work at Perimeter Institute has been supported by a Distinguished Visiting Research Chair position. The authors would like to thank the Aspen Center for Physics, which is supported by National Science Foundation grant No. PHY-2210452, where part of this work has been done.

\bibliography{etaW}
\bibliographystyle{jhep}

\appendix
\begin{widetext}
\section{Instanton induced potential}
\label{app:instanton}

As defined in the main text, the $SU(2)_L$ topological charge is
\begin{equation}
    Q =
    \frac{g^2}{32\pi^2}
    \int d^4x\,
    W^a_{\mu\nu}\widetilde W^{a\mu\nu}
    \in \mathbb Z .
\end{equation}
With the anomaly normalization used in the main text, a configuration with
$Q=\pm1$ carries the axion-dependent phase
\begin{equation}
    \exp\left[
        \pm i\left(
        \theta_W + N_f\frac{a}{f}
        \right)
    \right].
\end{equation}
In the renormalizable Standard Model this phase can be removed by an anomalous
$B+L$ rotation. It becomes physical once interactions that explicitly violate
$B+L$ are present, such as the $QQQL$ operator considered below
\cite{FileviezPerez:2014xju}.

For a constrained electroweak instanton of size $\rho$, the Euclidean action is
\begin{equation}
    S_E(\rho)
    =
    \frac{8\pi^2}{g^2(\mu)}
    +
    \pi^2\rho^2
    v^2
    +
    \mathcal O(\lambda \rho^4 v^4).
\end{equation}

\noindent The Higgs VEV, $v=246 \GeV$, gives the large-instanton suppression.

Following the standard one-instanton normalization of
Ref.~\cite{Morrissey:2005uza}, the bosonic instanton density can be written as
\begin{equation}
    d n_I(\rho)
    =
    C_2
    \left(
        \frac{8\pi^2}{g^2(\mu)}
    \right)^4
    e^{-8\pi^2/g^2(\mu)}
    (\mu\rho)^{b_0}
    \frac{d\rho}{\rho^5}
    e^{-\pi^2 v^2\rho^2}.
\end{equation}
Here $C_2$ is the one-instanton determinant coefficient given by Eq.(18) of \cite{Morrissey:2005uza}, and
\begin{equation}
    b_0=
    \frac{22}{3}
    -
    \frac{n_f}{3}
    -
    \frac{n_s}{6}.
\end{equation}
For the Standard Model,
\begin{equation}
    n_f=12,\qquad n_s=1,\qquad b_0=\frac{19}{6}.
\end{equation}
Equivalently, we define the one-loop invariant scale by
\begin{equation}
    \Lambda_{SU(2)}^{b_0}
    \equiv
    \mu^{b_0}
    e^{-8\pi^2/g^2(\mu)}
\end{equation}

The one-instanton background contains one fermion zero mode for each left-handed weak doublet. In the three-generation Standard Model there are twelve such zero modes, corresponding to
\begin{equation}
    \{Q_p^a,L_p\},
    \qquad
    p=1,2,3,
    \qquad
    a=1,2,3 .
\end{equation}
Therefore a contribution to the vacuum energy vanishes unless all twelve fermion zero modes are saturated. Thus, the instanton contribution to the axion potential is obtained by connecting the zero modes line as shown in Fig.~\ref{fig:inst}.

\begin{figure}
    \centering
    \includegraphics[width=0.3\linewidth]{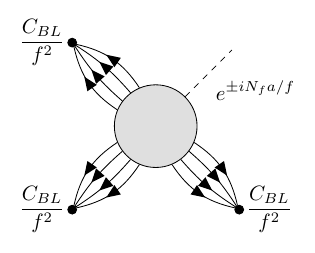}
    \caption{Leading instanton and anti-instanton contribution to the weak axion effective potential.}
    \label{fig:inst}
\end{figure}

In the rotated basis of Eq.~\eqref{eq:lagaxion}, the relevant interaction is
schematically
\begin{equation}
    \mathcal L_{QQQL}
    =
    \frac{C_{BL}}{f^2}QQQL
    +{\rm h.c.}
\end{equation}
The gauge, Lorentz, color, and flavor contractions are the usual ones for the dimension-six baryon-number-violating operator. For the estimate below, these details only affect an overall dimensionless coefficient. Three insertions of this operator are required to absorb the twelve zero modes. The relevant term in the Euclidean functional integral is
\begin{equation}
    \frac{1}{3!}
    \left[
    -
    \int d^4x\,
    \frac{C_{BL}}{f^2}QQQL
    \right]^3 .
\end{equation}

The zero-mode wavefunctions have the scaling form
\begin{equation}
    \psi_0(x-x_0;\rho)
    =
    \rho^{-2}
    \widehat\psi_0
    \left(
        \frac{x-x_0}{\rho}
    \right),
\end{equation}
up to corrections suppressed by $v\rho$. Thus an integrated local $QQQL$
overlap scales as
\begin{equation}
    \int d^4x\,QQQL[\psi_0]
    \sim
    \rho^{-4}.
\end{equation}
The three insertions therefore supply $\rho^{-12}/f^6$. This combines with the $\rho^{n_f/2}=\rho^6$ factor associated with the twelve fermionic zero-mode normalizations in the instanton measure. The net effect of the three local $QQQL$ insertions is therefore a factor of $(\rho f)^{-6}$. In the instanton calculation, the remaining finite spin, gauge-orientation, and flavor contractions are written as an order-one coefficient, $\kappa$ multiplying the Wilson coefficient,
\begin{equation}
    \kappa\frac{c_{BL}^3}{(4\pi)^6},
    \qquad
    c_{BL}\equiv |c_{BL}|e^{i\delta_{BL}} .
\end{equation}
For a general flavor tensor, $c_{BL}^3$ should be replaced by the corresponding product of three Wilson coefficients contracted with the zero-mode structure. We take $\kappa = 1$ for simplicity.

Adding the instanton and anti-instanton sectors gives the axion potential
\begin{equation}
    V_{\rm inst}(a)
    =
    -\Lambda_{\rm inst}^4
    \cos\left[
        N_f\frac{a}{f}
        +
        \theta_W
        +
        3\delta_{BL}
    \right].
\end{equation}
The factor of $3\delta_{BL}$ comes from the three insertions of the $QQQL$
operator. The amplitude is
\begin{align}
    \Lambda_{\rm inst}^4
    \simeq{}&
    2C_2
    \left(
        \frac{8\pi^2}{g^2(M_{\rm UV})}
    \right)^4
    \frac{|c_{BL}|^3}{(4\pi)^6}
    \mathcal I[M_{\rm UV}],
    \label{eq:weak-instanton-amplitude-app}
    \\
    \mathcal I[M_{\rm UV}]
    \equiv{}&
    \int_{M_{\rm UV}^{-1}}^\infty
    \frac{d\rho}{\rho^5}
    (\Lambda_{SU(2)}\rho)^{b_0}
    \frac{1}{(\rho f)^6}
    e^{-\pi^2 v^2\rho^2}.
    \label{eq:weak-instanton-integral-app}
\end{align}
Here $M_{\rm UV}$ is the scale at which the local $QQQL$ description should be
matched to its ultraviolet completion. The lower limit on the $\rho$ integral
reflects the fact that for $\rho\lesssim M_{\rm UV}^{-1}$ the instanton probes
the heavy physics that generated the effective operator.

The power of $\rho$ in Eq.~\eqref{eq:weak-instanton-integral-app} is
\begin{equation}
    -5+b_0-6
    =
    -5+\frac{19}{6}-6
    =
    -\frac{47}{6}.
\end{equation}
Thus the local EFT integral is dominated by the small-instanton region near
$\rho\simeq M_{\rm UV}^{-1}$, while the Higgs VEV regulates the large-$\rho$
region. For $M_{\rm UV}\gg v$, one finds parametrically
\begin{equation}
    \mathcal I[M_{\rm UV}]
    \simeq
    \frac{6}{41}
    \frac{\Lambda_{SU(2)}^{b_0}
    M_{\rm UV}^{41/6}}{f^6}
    \left[
        1+\mathcal O\left(\frac{v^2}{M_{\rm UV}^2}\right)
    \right].
\end{equation}
The precise finite coefficient depends on the ultraviolet completion of the
$QQQL$ operator, but the EFT fixes the zero-mode selection rule, the phase
structure, and the small-instanton power counting.

Taking $M_{\rm UV}\simeq f\simeq M_{\rm pl}$ gives the estimate quoted in the
main text,
\begin{equation}
    \Lambda_{\rm inst}^4
    \simeq
    \left(6\times10^{-6}\,{\rm eV}\right)^4
    |c_{BL}|^3 .
\end{equation}
Therefore, without modifying the Standard Model weak gauge sector, the
electroweak-instanton contribution is far below the observed dark-energy
density. In this setup it fixes the anomalous $B+L$ phase structure, but it
does not provide the dominant contribution to the weak axion potential.

\section{PMNS conventions}
\label{app:PMNS}

We relate the flavor and Majorana mass eigenstates through
\begin{align}
    \nu_\alpha = \sum_i V_{\alpha i}\,\nu_i,
\end{align}
where $\alpha=e,\mu,\tau$ and $i=1,2,3$.
The full PMNS matrix is
\begin{align}
    V =
    U\,{\rm diag}\!\left(
        1,e^{i\alpha_{21}/2},e^{i\alpha_{31}/2}
    \right),
\end{align}
where $\alpha_{21}$ and $\alpha_{31}$ are the Majorana
phases. The Dirac part $U$ is related by charged-lepton
rephasings to the standard parametrization
\begin{align}
    \widetilde U =
    \begin{pmatrix}
        c_{12}c_{13}
        & s_{12}c_{13}
        & s_{13}e^{-i\delta_{\rm CP}}
        \\
        -s_{12}c_{23}
        -c_{12}s_{23}s_{13}e^{i\delta_{\rm CP}}
        & c_{12}c_{23}
        -s_{12}s_{23}s_{13}e^{i\delta_{\rm CP}}
        & s_{23}c_{13}
        \\
        s_{12}s_{23}
        -c_{12}c_{23}s_{13}e^{i\delta_{\rm CP}}
        & -c_{12}s_{23}
        -s_{12}c_{23}s_{13}e^{i\delta_{\rm CP}}
        & c_{23}c_{13}
    \end{pmatrix},
\end{align}
with $s_{ij}\equiv\sin\theta_{ij}$ and
$c_{ij}\equiv\cos\theta_{ij}$.
The flavor-aligned examples use $U=\widetilde U$.
The charged-lepton rephasing used for the democratic
direction is specified in Appendix~\ref{app:democratic}.

For the NuFit benchmark used in the uncertainty estimate,
we adopt the NuFit-6.0 normal-ordering best fit including
Super-K atmospheric data,
\begin{align}
    \theta_{12}
    &= \left(33.68^{+0.73}_{-0.70}\right)^\circ,
    \\
    \theta_{23}
    &= \left(43.3^{+1.0}_{-0.8}\right)^\circ,
    \\
    \theta_{13}
    &= \left(8.56^{+0.11}_{-0.11}\right)^\circ,
    \\
    \delta_{\rm CP}
    &= \left(212^{+26}_{-41}\right)^\circ.
\end{align}
Defining $\Delta m_{ij}^2\equiv m_i^2-m_j^2$,
the corresponding mass-squared splittings are
\begin{align}
    \Delta m_{21}^2
    &= (7.49\pm0.19)\times10^{-5}\,\mathrm{eV}^2,
    \\
    \Delta m_{31}^2
    &= \left(2.513^{+0.021}_{-0.019}\right)
       \times10^{-3}\,\mathrm{eV}^2.
\end{align}

For inverted ordering, the same analysis gives
\begin{align}
    \theta_{12}
    &= \left(33.68^{+0.73}_{-0.70}\right)^\circ,
    \\
    \theta_{23}
    &= \left(47.9^{+0.7}_{-0.9}\right)^\circ,
    \\
    \theta_{13}
    &= \left(8.59^{+0.11}_{-0.11}\right)^\circ,
    \\
    \delta_{\rm CP}
    &= \left(274^{+22}_{-25}\right)^\circ,
    \\
    \Delta m_{21}^2
    &= (7.49\pm0.19)\times10^{-5}\,\mathrm{eV}^2,
    \\
    \Delta m_{32}^2
    &= (-2.484\pm0.020)\times10^{-3}\,\mathrm{eV}^2.
\end{align}
Here $\Delta m_{31}^2=\Delta m_{32}^2+\Delta m_{21}^2$.

\section{Democratic direction and additional flavor protection}
\label{app:democratic}

The democratic flavor-breaking direction,
$|n_\star\rangle=|d\rangle=(1,1,1)^T/\sqrt{3}$,
provides an example of how neutrino parameters can control the
scale of the dark-energy potential. Its small overlap with the
heaviest neutrino eigenstate suppresses the neutrino
Coleman--Weinberg contribution. However, the trace condition
in Eq.~\eqref{eq:trcond} does not eliminate the charged-lepton
contribution in Eq.~\eqref{eq:charged_lepton_potential}.
Additional protection could arise from an extended neutrino
mass sector or symmetry-enforced cancellations.
We do not provide such a mechanism here, so the following
results describe the neutrino-sector contribution subject
to this requirement.

The charged-lepton fields may be rephased without changing
any oscillation observable. We define the democratic direction
in a convention where the second column of the Dirac PMNS
matrix is real and positive. Starting from the standard
parametrization $\widetilde U$ in Appendix~\ref{app:PMNS},
we write
\begin{align}
    P_\ell &=
    {\rm diag}\!\left(
    \frac{\widetilde U_{e2}^\ast}{|\widetilde U_{e2}|},
    \frac{\widetilde U_{\mu2}^\ast}{|\widetilde U_{\mu2}|},
    \frac{\widetilde U_{\tau2}^\ast}{|\widetilde U_{\tau2}|}
    \right),
    &
    U &= P_\ell\widetilde U .
\end{align}
Thus $U_{\alpha2}=|\widetilde U_{\alpha2}|$, and the democratic
direction aligns with the second neutrino eigenstate in the
tribimaximal limit. The breaking spurion and neutrino mass
matrix are defined in this same flavor basis.

The overlaps are
\begin{align}
    q_i
    \equiv \langle d|\nu_i^\ast\rangle
    =
    \frac{1}{\sqrt{3}}
    \sum_{\alpha=e,\mu,\tau}V_{\alpha i}^\ast .
\end{align}
Using
$V=U\,{\rm diag}(1,e^{i\alpha_{21}/2},e^{i\alpha_{31}/2})$,
we obtain
\begin{align}
    q_i &=
    \frac{e^{-i\alpha_i/2}}{\sqrt{3}}
    \sum_{\alpha=e,\mu,\tau}
    \frac{\widetilde U_{\alpha2}}
         {|\widetilde U_{\alpha2}|}
    \widetilde U_{\alpha i}^\ast ,
    \qquad
    \alpha_i=(0,\alpha_{21},\alpha_{31}) .
    \label{eq:qi_rephased}
\end{align}
For the third mass eigenstate, defining
\begin{align}
    A_\mu &\equiv
    c_{12}c_{23}
    -s_{12}s_{23}s_{13}e^{i\delta_{\rm CP}},
    \\
    A_\tau &\equiv
    -c_{12}s_{23}
    -s_{12}c_{23}s_{13}e^{i\delta_{\rm CP}},
\end{align}
gives
\begin{align}
    q_3 =
    \frac{e^{-i\alpha_{31}/2}}{\sqrt{3}}
    \left[
    s_{13}e^{i\delta_{\rm CP}}
    +c_{13}s_{23}\frac{A_\mu}{|A_\mu|}
    +c_{13}c_{23}\frac{A_\tau}{|A_\tau|}
    \right].
    \label{eq:q3_rephased}
\end{align}
The Majorana phase $\alpha_{31}$ therefore drops out of
$|q_3|^2$.

To expose the suppression analytically, we expand around
tribimaximal mixing,
\begin{align}
    \theta_{12}
    &= \sin^{-1}(1/\sqrt{3})
       +\epsilon\,\delta\theta_{12},
    \\
    \theta_{23}
    &= \frac{\pi}{4}
       +\epsilon\,\delta\theta_{23},
    \\
    s_{13} &= \epsilon\,\delta\theta_{13},
\end{align}
where $\epsilon$ tracks departures from this limit.
At leading order,
\begin{align}
    q_3
    \simeq
    \frac{e^{-i\alpha_{31}/2}}{\sqrt{3}}
    \left[
        \sqrt{2}\sin\left(\theta_{23}-\frac{\pi}{4}\right)
        +s_{13}\cos\delta_{\rm CP}
    \right].
    \label{eq:q3_democratic}
\end{align}
More explicitly,
\begin{align}
    |q_3|^2
    &=
    \frac{\epsilon^2}{3}
    \left(
        \sqrt{2}\,\delta\theta_{23}
        +\delta\theta_{13}\cos\delta_{\rm CP}
    \right)^2
    +\mathcal O(\epsilon^4)
    \\
    &\simeq
    \frac{1}{3}
    \left(
        s_{23}-c_{23}
        +s_{13}\cos\delta_{\rm CP}
    \right)^2 .
\end{align}
The overlap vanishes in the tribimaximal limit.
Away from this limit, the atmospheric mixing angle and
Dirac phase control the leading suppression, and their
contributions can partially cancel.

For normal ordering with $m_1\simeq0$ and $|q_2|\simeq1$,
the leading mixed solar--atmospheric contribution gives
\begin{align}
    \Lambda_\nu^4
    \sim
    \frac{m_2m_{\rm atm}^3}{8\pi^2}|q_3|^2 .
\end{align}
Taking $m_2\simeq8.7\,\mathrm{meV}$,
$m_{\rm atm}\simeq50\,\mathrm{meV}$, and $s_{13}\simeq0.15$,
we obtain
\begin{align}
    \Lambda_\nu^{d,\mathrm{NO}}
    \sim
    2.3\,\mathrm{meV}
    \sqrt{
        \left|
        \frac{\cos\delta_{\rm CP}}{0.52}
        +
        \frac{\sin\left(\theta_{23}-\frac{\pi}{4}\right)}
             {0.055}
        \right|
    }.
    \label{eq:Lambdanu_value}
\end{align}
Near a cancellation of the leading overlap, subleading
terms must be retained using Eq.~\eqref{eq:VCW_rankone}.

Figure~\ref{fig:LambdanuDeltaCP} shows the leading-invariant
estimate in the $(\delta_{\rm CP},\sin^2\theta_{23})$ plane.
The suppressed region follows the small overlap of the
breaking direction with the heaviest mass eigenstate.
For the NO$\nu$A benchmark, we use
$\sin^2\theta_{23}=0.559$, corresponding to
$\theta_{23}=48.4^\circ$, and $\delta_{\rm CP}=0.89\pi$,
while keeping the remaining oscillation parameters at
the NuFit central values.

\begin{figure}[b]
    \centering
    \includegraphics[width=0.58\linewidth]
        {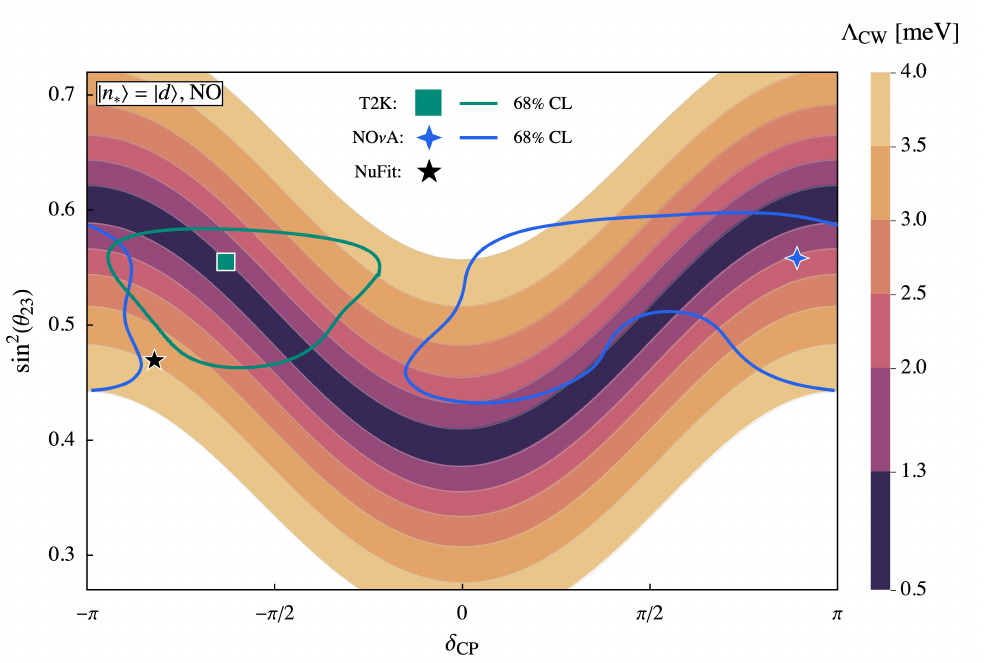}
    \caption{
        Neutrino-sector leading-invariant estimate of
        $\Lambda_{\rm CW}$ in the
        $(\delta_{\rm CP},\sin^2\theta_{23})$ plane for the
        democratic flavor-breaking direction
        $|n_\star\rangle=|d\rangle$.
        The colored contours show the potential scale in meV.
        The green square and contour show the normal-ordering
        T2K best-fit point and its $68\%$ CL region, while the
        blue four-pointed star and contour show the NO$\nu$A
        result. Additional protection against charged-lepton
        contributions is required for this estimate to
        determine the total potential.
    }
    \label{fig:LambdanuDeltaCP}
\end{figure}

The strong dependence on $\delta_{\rm CP}$ and $\theta_{23}$
connects this contribution to measurements of CP violation
and atmospheric mixing. Improved determinations by DUNE
and Hyper-Kamiokande
\cite{DUNE:2020jqi,Hyper-Kamiokande:2018ofw}
could therefore sharpen the predicted neutrino-sector
potential scale.

\end{widetext}

\end{document}